\documentclass[preprint,12pt]{elsarticle}
\usepackage{amsmath}
\usepackage[all]{xy}
\usepackage{color}
\usepackage{tikz}
\usepackage{extarrows}
\newtheorem{thm}{Theorem}
 \newtheorem{cor}{Corollary}
 \newtheorem{lem}{Lemma}
 \newtheorem{prop}{Proposition}

\begin{document}
\begin{frontmatter}
\title{B\"acklund transformation of the Geng-Xue system}

\author{Lihua Wu, Nianhua Li\footnote{Corresponding author. linianh@hqu.edu.cn}}

\address{School of Mathematical Sciences, Huaqiao University, Quanzhou, 362021, P. R. China.\\[5pt]
}

\begin{abstract}
We construct a B\"acklund transformation for the Geng-Xue system with the help of reciprocal and gauge transformations. Furthermore, we derive $N$-B\"acklund transformation for the Geng-Xue system resorting to Bianchi's permutability. As an application, we obtain some exact solutions of the Geng-Xue system including multi-kink, bell-shaped soliton. Finally, we discuss B\"acklund transformations for the Degasperis-Procesi and the Novikov equations, which are two reductions of the Geng-Xue system.
\bigskip

\end{abstract}

\end{frontmatter}
\section{Introduction}

The Camassa-Holm (CH) equation \cite{Holm}
\begin{equation}
m_t+u m_x+2u_x m=0, \ \ m=u-u_{xx},
\end{equation}
arises as a model for long waves in shallow water by the asymptotic
approximation of Hamiltonian for Euler's equations. It is a completely integrable system since it has Lax pair with bi-Hamiltonian structure, and may be solved by the B\"acklund transformation \cite{Rasin} as well as the inverse scattering transformation  \cite{Constantin,Gerdjikov}. The CH equation can be linked to the first negative flow of the KdV hierarchy by a reciprocal transformation \cite{Fuchssteiner}. One important feature for the CH equation is admiting peakon solutions
\cite{Szmigielski,Sattinger,Beals}, which have discontinuities in $x$-derivative but both one-sided derivatives exist and differ only by a sign at the crest. Henceforth, integrable equations with peakon solutions have attracted much attention in recent years \cite{Popowicz}.

The Geng-Xue (GX) system \cite{Geng}
 \begin{eqnarray}\label{2novikov}
 \begin{aligned}
 &m_{t}+3u_{x}vm+uvm_{x} =0, \quad \quad m=u-u_{xx},\\
&n_{t}+3v_{x}un+uvn_{x} =0,  \quad \quad \quad n=v-v_{xx}, \end{aligned}
\end{eqnarray}
is a coupled integrable CH type system with cubic nonlinearity and admits a Lax pair and associated bi-Hamiltonian structure \cite{Liuq}. It is reciprocally connected with a first negative flow of a modified Boussinesq hierarchy \cite{Li}. Lundmark and Szmigielski throughly studied inverse spectral problem and got multi-peakon solutions of the GX system \cite{ldk}.
 Very recently, multi-kink solutions of the GX system were obtained by Darboux transformation \cite{Liliu}.

In addition, the GX system is closely related to
the Degasperis-Procesi (DP) equation \cite{Degasperis}
\begin{equation}\label{dp}
m_t+um_x+3u_xm=0,\quad m=u-u_{xx},
\end{equation}
and
the Novikov equation \cite{Novikov}
\begin{equation}\label{novikov}
m_{t}+u^{2}m_{x}+3uu_{x}m=0,\quad m=u-u_{xx},
\end{equation}
since they can be reduced from \eqref{2novikov} as $v=1$ and $v=u$, respectively.
There are many works on their Lax representations, bi-Hamiltonian structures, reciprocal partners and exact solutions \cite{wang1}-\cite{Szmigielski3}.

 B\"acklund transformations (BTs), originated from the differential geometry,  play an important role in the theory of integrable systems, such as searching exact solutions, integrable discretization, as well as constructing symmetries, etc. \cite{b,la,rs}. However, in view of the speciality of the spectral problem for the CH type equations, it is hard to construct their BTss directly. Recently, Rasin and Schiff discussed BT for the CH equation with the help of reciprocal transformation and concluded that it involves not only the dependent variables but also the independent spatial variables \cite{Rasin}. Later on Mao, Liu et al construct BTs for the DP, the Novikov and the short pulse equations \cite{mw, mh, ml,Huang}. As far as we know, there is no results on the BT of the GX system. The aim of this paper is to construct the $N$-BT of the GX system.

The paper is arranged as follows. In section 2, we first introduce a reciprocal
transformation to relate the GX system with an associated GX (aGX) system, and further to a negative flow of the Boussinesq hierarchy by a gauge transformation. With the aid of these two transformations, we get a BT for the GX system from the Darboux transformation of the negative Boussinesq flow.
In section 3, using the Bianchi's permutability, we derive $2$ and $N$-BT for the GX system.
In section 4, we apply BT to obtain exact solutions for the GX system such as multi-kink, bell-shaped soliton etc..
In section 5, the BT for the DP equation and the Novikov equation are discussed.

\section{B\"acklund transformation of the Geng-Xue system}
According to Ref. \cite{Geng}, the GX system \eqref{2novikov} admits the Lax pair
\begin{equation}\label{gxlax}
\psi_x=U\psi,\quad \quad \quad \psi_t=V\psi,
\end{equation}where $\psi=(\psi_1,\psi_2,\psi_3)^{T}$ and
\begin{equation*}
U=\left[
\begin{matrix}
                   0 & \lambda m & 1\\
                   0 & 0 & \lambda n \\
                   1 & 0 & 0 \\
                  \end{matrix}
                \right],\quad V=\left[
\begin{matrix}
                  -u_xv & \frac{u_x}{\lambda}-\lambda uvm  & u_xv_x\\
                   \frac{v}{\lambda} & -\frac{1}{\lambda^2}+u_xv-uv_x & -\lambda uvn-\frac{v_x}{\lambda} \\
                   -uv &  \frac{u}{\lambda} & uv_x \\
                  \end{matrix}
                \right].
\end{equation*}
It was shown that the GX system has infinitely many conservation laws \cite{Geng,Li} in which the first one is
$$q_t=(-uvq)_x,\quad\quad\quad q=(mn)^{\frac{1}{3}}.$$
This naturally defines a reciprocal transformation
\begin{equation}\label{gxre}
dy=qdx-uvqdt,\quad \quad \quad d\tau=dt.
\end{equation}
Applying \eqref{gxre} to the Lax pair (\ref{gxlax}), we have
\begin{equation}\label{agxlax}
\psi_y=F\psi,\quad \quad \quad \psi_\tau=G\psi,
\end{equation}
where
\begin{equation*}
F=\left[
\begin{matrix}
                  0 & \lambda p & \frac{1}{q}\\
                  0 & 0 & \lambda \frac{q}{p} \\
                  \frac{1}{q} & 0 & 0 \\
                  \end{matrix}
                \right],\quad G=\left[
\begin{matrix}
                -u_yvq & \frac{u_yq}{\lambda}  & uv+u_yv_yq^2\\
                   \frac{v}{\lambda} & u_yvq-uv_yq-\frac{1}{\lambda^2} & -\frac{v_yq}{\lambda} \\
                  0 &   \frac{u}{\lambda} & uv_yq \\
                  \end{matrix}
                \right],
\end{equation*}and $p=\frac{m}{q}$. Direct calculation shows that the compatibility condition of linear system (\ref{agxlax}) yields the aGX system
\begin{eqnarray}\label{agxsystem}
 \begin{aligned}
 &p_\tau=pq(uv_y-2u_yv),\quad \quad \quad u_{yy}q^2+u_yqq_y+pq-u=0, \\
& q_\tau=-q^2(uv)_y,\quad \quad \quad \quad \quad \ \ v_{yy}q^2+qq_yv_y+p^{-1}q^2-v=0. \end{aligned}
\end{eqnarray}
Eliminating $\psi_1,\psi_2$ from (\ref{agxlax}), we obtain a scalar spectral problem for the wave function $\psi_3$. Under a gauge transformation $\psi_3=p^{\frac{1}{3}}q^{-\frac{2}{3}}\phi$,  the scalar spectral problem is converted to the classical spectral problem of the Boussinesq hierarchy
 \begin{equation}\label{Boussinesq}
(\partial_y^3+Q_1\partial_y+Q_2)\phi=(\partial_y-r)(\partial_y-s)(\partial_y+r+s)\phi=\lambda^2\phi,
 \end{equation}
where
\begin{equation}\label{gxpo}
r=\frac{2p_y}{3p}-\frac{q_y}{3q},\quad \quad \quad s=-\frac{p_y}{3p}-\frac{q_y}{3q}-\frac{1}{q}.
\end{equation}
With the aid of the classical DT of the Boussinesq hierarchy \cite{Leble}, we get a DT for the aGX system (\ref{agxsystem}).
\begin{prop}
The Lax presentation (\ref{agxlax}) is covariant under the DT:
\begin{eqnarray}\label{DTgx}
 \begin{aligned}&\psi_{[1]}=T(\lambda_1,a_1,b_1)\psi,\quad \quad \quad   T(\lambda_1,a_1,b_1)=\left[
\begin{matrix}
                   -\frac{a_{1}}{c_{1}}& \frac{\lambda (a_1^2-1) }{\lambda_1 b_1c_1}  & \frac{1}{c_1}\\
                  0& -1 & \frac{\lambda b_1}{\lambda_1} \\
                    \frac{1}{c_1} & 0 & -\frac{a_1}{c_1} \\
                  \end{matrix}
                \right], \\
&p_{[1]}=\frac{q(a_1^2-1)}{pb_1^2c_1},\quad \quad \quad  \ \ q_{[1]}=\frac{a_1^2-1}{\lambda_1pb_1},\\
&u_{[1]}=\frac{1}{c_1}(ua_1-u_yq),\quad \quad \quad   \ \ v_{[1]}=\frac{c_1}{a_1^2-1}(va_1-v_yq-\frac{b_1}{\lambda_1}), \end{aligned}
\end{eqnarray}where $a_1=\frac{\varphi_1}{\varphi_3}, b_1=\frac{\varphi_2}{\varphi_3}$, $c_1=\sqrt{|a_1^2-1|}$, and $(\varphi_1,\varphi_2,\varphi_3)^{T}$ is a special solution of (\ref{agxlax}) or (\ref{gxlax}) at $\lambda=\lambda_1$.
\end{prop}

To construct a BT for the GX system, it is important to observe that
\begin{equation}\label{q1}
\frac{1}{q_{[1]}}=\frac{1}{q}+\frac{a_{1,y}}{a_1^2-1},\quad \quad \quad u_{[1]}v_{[1]}=uv+\frac{a_{1,\tau}}{a_1^2-1}.
\end{equation}
Taking (\ref{gxre}) and \eqref{q1} into account, we arrive at
$$dx_{[1]}=\frac{1}{q_{[1]}}dy+u_{[1]}v_{[1]}d\tau =d(x-\frac{1}{2}{\rm ln} \lvert \frac{a_1+1}{a_1-1}\rvert).$$
Integrating on both sides of this equation and choosing the integration constant to be zero, we obtain
\begin{equation}\label{x1}
x_{[1]}=x-\frac{1}{2}{\rm ln} \lvert \frac{a_1+1}{a_1-1}\rvert.
\end{equation}
Given these preparations, the following proposition holds.

\begin{prop}
The GX system admits a BT
\begin{eqnarray}
 \begin{aligned}&x_{[1]}=x-\frac{1}{2}{\rm ln} \lvert \frac{a_1+1}{a_1-1}\rvert,\quad \quad \quad \quad t_{[1]}=t, \\
&u_{[1]}=\frac{1}{c_1}(ua_1-u_x),\\
&v_{[1]}=\frac{c_1}{a_1^2-1}(va_1-v_x-\frac{a_{1,x}+a_1^2-1}{\lambda_1^2m}), \end{aligned}
\end{eqnarray}where $c_1=\sqrt{|a_1^2-1|}$, and $a_1$ is controlled by the system
\begin{eqnarray}
\begin{aligned}&a_{1,xx}=(\frac{m_x}{m}-a_1)(a_{1x}+a_1^2-1)-2a_1a_{1x}+\lambda_1^2mn, \\
&a_{1,t}=\frac{u_x-ua_1}{\lambda_1^2m}(a_{1,x}+a_1^2-1)-(uva_1)_x+uv+u_xv_x.\end{aligned}
\end{eqnarray}
\end{prop}

\section{$N$-B\"acklund transformation of the Geng-Xue system}
In this section, we shall first deduce a 2-BT for the GX system, and then extend it to $N$-BT. To begin with, let us show the diagram of Bianchi's permutability as follows.
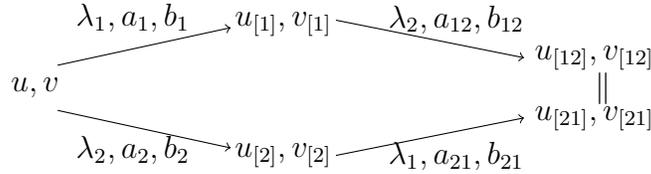
\begin{figure}[htb]
\centering\begin{tikzpicture}
\draw [->](0.3,0.3)--(2.6,0.8);
\draw [->](0.3,-0.3)--(2.6,-0.8);
\draw [->](4.0,-0.9)--(6.5,-0.4)  node[right] {$u_{[21]},v_{[21]}$} ;
\draw [->](4.0,0.9)--(6.5,0.4) node[right] {$u_{[12]},v_{[12]}$} ;
\node at ( 0,0) {$u,v $};
\node at ( 7.3,0) [right]{$\parallel$};
\node at ( 3.3,0.85)  {$u_{[1]},v_{[1]}$};
\node at ( 3.3,-0.9)  {$u_{[2]},v_{[2]}$};
\node at ( 1.3,0.60) [above] {$\lambda_1, a_1, b_1$};
\node at ( 1.3,-0.50) [below] {$\lambda_2,  a_2, b_2$};
\node at ( 5.6,0.58) [above] {$\lambda_2, a_{12}, b_{12}$};
\node at ( 5.6,-0.58) [below] {$\lambda_1, a_{21}, b_{21}$};
\end{tikzpicture}
\caption{Bianchi permutability}\label{gxBianchi}
\end{figure}

\noindent Using this Bianchi's permutability, we have
  \begin{eqnarray}
T(\lambda_2,a_{12},b_{12})T(\lambda_1,a_1,b_1)=T(\lambda_1,a_{21},b_{21})T(\lambda_2,a_2,b_2),
\end{eqnarray}
which leads to
\begin{eqnarray*}\label{ab12}
&&\hspace{-0.6cm}a_{12}=\frac{\lambda_2b_2(a_1^2-1)+\lambda_1b_1(1-a_1a_2)}{\lambda_1b_1(a_2-a_1)}, \ \ a_{21}=\frac{\lambda_1b_1(a_2^2-1)+\lambda_2b_2(1-a_1a_2)}{\lambda_2b_2(a_1-a_2)}, \\ &&\hspace{-0.6cm}b_{12}=\frac{(\lambda_2b_1-\lambda_1b_2)c_1}{\lambda_1(a_2-a_1)},\quad \quad \quad
 b_{21}=\frac{\lambda_1c_2}{\lambda_2c_1}b_{12},\quad \quad \quad c_{21}=\frac{(a_2^2-1)\lambda_1b_1c_1}{(a_1^2-1)\lambda_2b_2c_2}c_{12}.
\end{eqnarray*}
Then, based on the Proposition 2, we have 2-BT for the GX system. The main result is stated as follows.
\begin{prop}
The GX system admits a 2-BT
\begin{eqnarray}
 \begin{aligned}
 &x_{[12]}=x-\frac{1}{2}{\rm ln} \lvert \frac{(a_1+1)(a_{12}+1)}{(a_1-1)(a_{12}-1)}\rvert,\quad \quad \quad \quad \quad  t_{[12]}=t, \\
 &u_{[12]}=\frac{1}{c_1c_{12}}[u(a_1a_{12}+1)-u_x(a_1+a_{12})-\frac{a_1^2-1}{\lambda_1b_1}],\\
 &v_{[12]}=\frac{c_1c_{12}}{(a_1^2-1)(a_{12}^2-1)}[v(a_1a_{12}+1)-(v_x+\frac{b_1}{\lambda_1})(a_1+a_{12})]\\
& \quad\quad\quad -\frac{c_1c_{12}}{(a_2-a_1)(a_{12}^2-1)}(\frac{b_1}{\lambda_1}-\frac{b_2}{\lambda_2}).
\end{aligned}
\end{eqnarray}
Here $c_{12}=\sqrt{\lvert a_{12}^2-1\rvert}$, $a_2=\frac{h_1}{h_3}, b_2=\frac{h_2}{h_3}$, and $(h_1,h_2,h_3)^{T}$ is a special solution of (\ref{gxlax}) at $\lambda=\lambda_2$.
\end{prop}

Next, we will derive $N$-BT of the GX system. For convenience, let us denote natural permutation from $1$ to any positive integer $N$ by $\widehat{N}$, i.e. $\widehat{N}=12\cdots N$. Then, constructing the $N$-BT comes down to give compact forms for $x_{[\widehat{N}]},a_{[\widehat{N}]},u_{[\widehat{N}]}$ and $v_{[\widehat{N}]}$.
In fact, it follows from Proposition 1 that
\begin{equation}\label{expA}
x_{[\widehat{N}]}=x-\frac{1}{2}{\rm ln}| \frac{(a_1+1)(a_{12}+1)...(a_{\widehat{N}}+1)}{(a_1-1)(a_{12}-1)...(a_{\widehat{N}}-1)}|.
\end{equation}
Since it's not easy to obtain compact form for $a_{[\widehat{N}]}$ directly, we
define $w_{\widehat{N}}$ by
\begin{equation}
 \frac{(a_1+1)(a_{12}+1)...(a_{\widehat{N}}+1)}{(a_1-1)(a_{12}-1)...(a_{\widehat{N}}-1)}=\frac{w_{\widehat{N}}+1}{w_{\widehat{N}}-1},
\end{equation}
which implies that
\begin{equation}\label{an2}
a_{\widehat{N}}=\frac{1-w_{\widehat{N-1}}w_{\widehat{N}}}{w_{\widehat{N}}-w_{\widehat{N-1}}}.
\end{equation}
We first devote ourselves to arriving at a recurrence relation for $w_{\widehat{N}}$ to obtain its expression of compact form, and hence for that of $a_{\widehat{N}}, \ x_{\widehat{N}}, \ u_{\widehat{N}},\ v_{\widehat{N}}$.

Resorting to Bianchi's permutability, we get the recurrence relations
\begin{eqnarray}\label{an1}
&&a_{\widehat{N}} = \frac{(a_{\widehat{N-1}}^2-1)\lambda_Nb_{\widehat{N-2}N}+\lambda_{N-1}b_{\widehat{N-1}}(1-a_{\widehat{N-2}N}a_{\widehat{N-1}})
}{\lambda_{N-1}b_{\widehat{N-1}}(a_{\widehat{N-2}N}-a_{\widehat{N-1}})},\\   \label{bn}
&&b_{\widehat{N}}=\frac{\lambda_Nb_{\widehat{N-1}}-\lambda_{N-1}b_{\widehat{N-2}N}}{\lambda_{N-1}(a_{\widehat{N-2}N}-a_{\widehat{N-1}})}c_{\widehat{N-1}},
\end{eqnarray}
where $c_{\widehat{N}}=\sqrt{|a_{\widehat{N}}^2-1|}$. Moreover,  introducing
\begin{eqnarray}\label{sin1}
\sigma^{N}_{\widehat{N-1}}=\frac{b_{\widehat{N-2}N}}{b_{\widehat{N-1}}},\quad \quad \quad  N\geq2,
\end{eqnarray}and inserting (\ref{an2}) into  (\ref{an1}), one infers
\begin{equation}\label{wn2}
w_{\widehat{N}}=\frac{\lambda_N\sigma^N_{\widehat{N-1}}w_{\widehat{N-1}}(w_{\widehat{N-2}N}-w_{\widehat{N-2}})
   +\lambda_{N-1}w_{\widehat{N-2}N}(w_{\widehat{N-2}}-w_{\widehat{N-1}})}
   {\lambda_N\sigma^N_{\widehat{N-1}}(w_{\widehat{N-2}s}-w_{\widehat{N-2}})
   +\lambda_{N-1}(w_{\widehat{N-2}}-w_{\widehat{N-1}})},
\end{equation}or equivalently
\begin{equation}\label{sin2}
\sigma^{N}_{\widehat{N-1}}=\frac{\lambda_{N-1}}{\lambda_{N}}\frac{(w_{\widehat{N-2}}-w_{\widehat{N-1}})(w_{\widehat{N-2}N}-w_{\widehat{N}})}{(w_{\widehat{N-2}}
-w_{\widehat{N-2}N})(w_{\widehat{N-1}}-w_{\widehat{N}})}.
\end{equation}

 We are now in a position to obtain determinant expressions for $w_{\widehat{N}}$ and $\sigma^{N}_{\widehat{N-1}}$ . A natural idea is to guess their expressions by observation of explicit formulae for $N\leq3$ and then prove them. In fact, it is not hard to show that
 the first several members in \eqref{wn2} and  \eqref{sin2} are
\begin{equation}\label{w13}
 \begin{aligned}
&w_1 = a_1, \quad\quad\quad
  w_{12} = \frac{\lambda_1b_1a_2-\lambda_2b_2a_1}{\lambda_1b_1-\lambda_2b_2},  \\
&w_{123}=  \frac{\lambda_1b_1(a_3\lambda_2^2-a_2\lambda_3^2)+\lambda_2b_2(a_1\lambda_3^2-a_3\lambda_1^2)
  +\lambda_3b_3(a_2\lambda_1^2-a_1\lambda_2^2)}{\lambda_1b_1(\lambda_2^2-\lambda_3^2)+\lambda_2b_2(\lambda_3^2-\lambda_1^2)
  +\lambda_3b_3(\lambda_1^2-\lambda_2^2)},\\
& \sigma^{2}_{1} =\frac{b_{2}}{b_{1}}, \quad\quad\quad
\sigma^{3}_{12} = \frac{b_{13}}{b_{12}}=\frac{(\lambda_3b_1-\lambda_1b_3)(a_2-a_1)}{(\lambda_2b_1-\lambda_1b_2)(a_3-a_1)}.
  \end{aligned}
  \end{equation}
 In view of \eqref{w13}, we introduce the following determinant
\begin{equation}
\begin{aligned}
 \Delta_N=\left\{\begin{array}{lll}
\left|\begin{matrix}
1&a_1 & \lambda_1b_1  &\cdots & \lambda_1^{2k} \\
\vdots& \vdots & \vdots &  & \vdots\\
1&a_N & \lambda_Nb_N & \cdots & \lambda_N^{2k}
\end{matrix}
\right|,  \ \ N=3k+1,\\\\

\left|\begin{matrix}
1&a_1 & \lambda_1b_1  & \cdots & \lambda_1^{2k}a_1 \\
\vdots& \vdots & \vdots &  & \vdots\\
1&a_N & \lambda_Nb_N  &\cdots & \lambda_N^{2k}a_N
\end{matrix}\right|,  \ \ N=3k+2, \\\\

\left|\begin{matrix}
1&a_1 & \lambda_1b_1  & \cdots & \lambda_1^{2k+1}b_1 \\
\vdots& \vdots & \vdots &  & \vdots\\
1&a_N& \lambda_Nb_N & \cdots & \lambda_N^{2k+1}b_N
\end{matrix}\right|,  \ \ N=3k+3,
\end{array}
\right.
\end{aligned}
\end{equation} for $k\in \mathbf{N}$.

\begin{thm}
The expressions for $\sigma^{N}_{\widehat{N-1}}$ and $w_{\widehat{N}}$  in terms of determinant $\Delta_N$ read
\begin{eqnarray}\label{wn3}
&&\hspace{-1.2cm}w_{\widehat{N}}=\frac{A_N}{B_N},\ \ N\geq1,\\ \label{sin3}
&&\hspace{-1.2cm}\sigma^{N}_{\widehat{N-1}}=\frac{\lambda_{N-1}(A_{N-2}B_{N-1}-A_{N-1}B_{N-2})(C_{N-1}B_{N}-A_{N}D_{N-1})}
{\lambda_{N}(A_{N-2}D_{N-1}-C_{N-1}B_{N-2})(A_{N-1}B_{N}-A_{N}B_{N-1})},  N\geq3,
\end{eqnarray}
 where
\begin{eqnarray*}
&&A_N=\Delta_{N+1}\left[\begin{matrix}
N+1\\ 1
\end{matrix}\right],\quad \quad \quad \quad   C_{N-1}=\Delta_{N}\left[\begin{matrix}
N-1\\
1
\end{matrix}\right],\\
&& B_N=\Delta_{N+1}\left[\begin{matrix}
N+1\\ 2
\end{matrix}\right],\quad \quad \quad \quad  D_{N-1}=\Delta_{N}\left[\begin{matrix}
N-1\\
2
\end{matrix}\right].
\end{eqnarray*}
\end{thm}
Here $J\left[
         \begin{array}{cccc}
           i_1 &i_2& \cdots&i_k  \\
           j_1& j_2& \cdots&j_k\\
         \end{array}
       \right]$
denotes the determinant by removing $i_1, \cdots, i_k$ rows and $j_1, \cdots, j_k$ columns from the determinant $J$.

To prove the theorem, we need two useful identities displayed in the following Lemma.
\begin{lem} Assume that $\pi$ is a $(N+2)\times N$ matrix, $\chi_k$ are $N+2$ order column vectors. Then we have
\noindent \begin{enumerate}
\item The Pl\"ucker relation\begin{equation*}
|\pi,\chi_1,\chi_2||\pi,\chi_3,\chi_4|-|\pi,\chi_1,\chi_3|
|\pi,\chi_2,\chi_4|+|\pi,\chi_1,\chi_4|
|\pi,\chi_2,\chi_3|=0.
\end{equation*}
\item The Jacobi identity
\begin{equation*}
\begin{array}{lll}\label{jco}
J \times J\left[ \begin{array}{cccc}
           i_1 &i_2  \\
           j_1& j_2
         \end{array}
       \right]=J\left[
         \begin{array}{cccc}
           i_1  \\
           j_1
         \end{array}
       \right]\times J\left[
         \begin{array}{cccc}
           i_2  \\
           j_2
         \end{array}
       \right]-J\left[
         \begin{array}{cccc}
           i_1  \\
           j_2
         \end{array}
       \right]\times J\left[
         \begin{array}{cccc}
           i_2  \\
           j_1
         \end{array}
       \right].
\end{array}
\end{equation*}
\end{enumerate}
\end{lem}

\noindent{\bf Proof of Theorem 1:} Here we only prove the case of $N=3k+2$ by the method of mathematical induction, because the other two cases can be verified similarly.

First, it is easy to check that both \eqref{wn3} and \eqref{sin3} are true for $N\leq3$. Next, assume \eqref{wn3} and \eqref{sin3} hold for $N-1$, our task is to verify them for $N$. In view of \eqref{bn} and \eqref{sin1}, it is straightforward to know that
\begin{eqnarray}\label{sin+1}
&&\sigma^{N+1}_{\widehat{N}}=\frac{b_{\widehat{N-1} N+1}}{b_{\widehat{N}}}=\frac{a_{\widehat{N-2}N}
-a_{\widehat{N-1}}}{a_{\widehat{N-2}N+1}-a_{\widehat{N-1}}}\frac{\lambda_{N+1}-\lambda_{N-1}\sigma^{N+1}_{\widehat{N-1}}}{\lambda_N-\lambda_{N-1}\sigma^{N}_{\widehat{N-1}}}.
\end{eqnarray}
On the one hand, it follows from (\ref{an2}) that
\begin{eqnarray}\label{th2}
&&\hspace{-1.5cm}\frac{a_{\widehat{N-2}N}-a_{\widehat{N-1}}}{a_{\widehat{N-2}N+1}-a_{\widehat{N-1}}}=\frac{(w_{\widehat{N-2}N}
-w_{\widehat{N-1}})(w_{\widehat{N-2}}-w_{\widehat{N-2}N+1})}{(w_{\widehat{N-2}N+1}-w_{\widehat{N-1}})(w_{\widehat{N-2}}-w_{\widehat{N-2}N})}\nonumber\\
&&\hspace{1.6cm}=\frac{(C_{N-1}B_{N-1}-A_{N-1}D_{N-1})(A_{N-2}H_{N-1}-B_{N-2}E_{N-1})}
{(E_{N-1}B_{N-1}-A_{N-1}H_{N-1})(A_{N-2}D_{N-1}-C_{N-1}B_{N-2})}
\end{eqnarray}with
\begin{equation*}
 E_{N-1}=\Delta_{N+1}\left[
         \begin{array}{ccc}
           N-1 &N  \\
           1&N+1\\
         \end{array}
       \right], \quad
      H_{N-1}=\Delta_{N+1}\left[
         \begin{array}{ccc}
           N-1 &N \\
           2&N+1 \\
         \end{array}
       \right].
\end{equation*}
On the other hand, by inductive hypotheses, one infers
\begin{equation}
\begin{aligned}\label{th3}
\frac{\lambda_{N+1}-\lambda_{N-1}\sigma^{N+1}_{\widehat{N-1}}}{\lambda_N-\lambda_{N-1}\sigma^{N}_{\widehat{N-1}}}=
\frac{\lambda_NR_1}{\lambda_{N+1}R_2}\frac{(A_{N-2}D_{N-1}-B_{N-2}C_{N-1})(A_{N-1}B_N-A_NB_{N-1})}
{(A_{N-2}H_{N-1}-B_{N-2}E_{N-1})(A_{N-1}D_{N}-B_{N-1}C_{N})},
\end{aligned}
\end{equation}
where
\begin{eqnarray*}
&&R_1=\lambda^2_{N+1}(A_{N-2}H_{N-1}-B_{N-2}E_{N-1})(A_{N-1}D_N-C_NB_{N-1})\\
&&\hspace{0.95cm}-\lambda^2_{N-1}(A_{N-2}B_{N-1}-A_{N-1}B_{N-2})(E_{N-1}D_{N}-C_{N}H_{N-1}),\\
&&R_2=\lambda^2_{N}(A_{N-2}D_{N-1}-B_{N-2}C_{N-1})(A_{N-1}B_N-A_NB_{N-1})\\
&&\hspace{0.95cm}-\lambda^2_{N-1}(C_{N-1}B_{N}-A_{N}D_{N-1})(A_{N-2}B_{N-1}-A_{N-1}B_{N-2}).
\end{eqnarray*}
Substituting \eqref{th2} and \eqref{th3} into \eqref{sin+1}, one has
\begin{equation}
\begin{aligned}\label{sin5}
&\sigma^{N+1}_{\widehat{N}}=\frac{\lambda_N(A_{N-1}B_N-A_NB_{N-1})}{\lambda_{N+1}(A_{N-1}D_{N}-B_{N-1}C_{N})}\frac{R_1(C_{N-1}B_{N-1}-A_{N-1}D_{N-1})}
{R_2(E_{N-1}B_{N-1}-A_{N-1}H_{N-1})}.
\end{aligned}
\end{equation}
Before proceeding further, let us list some useful identities obtained from the Jacobi identity and Pl$\ddot{u}$cker relation as follows
\begin{eqnarray}\label{jac}
 \begin{aligned}
 &C_{N-1}B_{N-1}-A_{N-1}D_{N-1}=\Delta_N \Delta_{N}\left[\begin{array}{cc}
           N-1&N \\ 1&2
         \end{array}
       \right],\\
&E_{N-1}B_{N-1}-A_{N-1}H_{N-1}=\Delta_{N+1}\left[\begin{array}{c}
           N  \\ N+1
         \end{array}
       \right] \Delta_{N}\left[\begin{array}{cc}
           N-1&N  \\ 1&2
         \end{array}
       \right],\end{aligned}
\end{eqnarray}
\begin{eqnarray}\label{plu2}
 \begin{aligned}
 &A_{N-1}B_N-A_NB_{N-1}=\Delta_{N} \Delta_{N+1}\left[\begin{array}{cc}
           N&N+1  \\ 1&2
         \end{array}
       \right],  \\
 &C_{N-1}B_{N}-A_{N}D_{N-1}=\Delta_{N} \Delta_{N+1}\left[\begin{array}{cc}
           N-1 &N+1 \\ 1&2
         \end{array}
       \right],\\
 &A_{N-1}D_N-C_NB_{N-1}=\Delta_{N+1}\left[\begin{array}{cc}
           N  \\  N+1
         \end{array}
       \right] \Delta_{N+1}\left[\begin{array}{cc}
           N&N+1 \\ 1&2
         \end{array}
       \right], \\
 &E_{N-1}D_{N}-C_{N}H_{N-1}= \Delta_{N+1}\left[\begin{array}{cc}
           N  \\  N+1
         \end{array}
       \right] \Delta_{N+1}\left[\begin{array}{cc}
           N-1&N  \\ 1&2
         \end{array}
       \right],\\
       &A_{N-2}H_{N-1}-B_{N-2}E_{N-1}=\Delta_{N+1}\left[\begin{array}{cc}
           N-1&N  \\ N& N+1
         \end{array}
       \right] \Delta_{N}\left[\begin{array}{cc}
           N-1&N  \\ 1&2
         \end{array}
       \right],
 \end{aligned}
\end{eqnarray}
whose proofs will be given in appendix.
With the help of \eqref{jac} and \eqref{plu2}, a direct calculation gives rise to
\begin{equation}\label{g3g4}
\sigma^{N+1}_{\widehat{N}}=\frac{\lambda_N(A_{N-1}B_N-A_NB_{N-1})}{\lambda_{N+1}(A_{N-1}D_N-B_{N-1}C_N)}\frac{R_3}{R_4},
\end{equation}where
\begin{eqnarray*}
&&R_3=\lambda^2_{N+1}\Delta_{N+1}\left[\begin{array}{cc}
           N-1& N \\ N&  N+1
         \end{array}
       \right] \Delta_{N+1}\left[\begin{array}{cc}
           N &N+1  \\ 1&2
         \end{array}
       \right]\\
&&\hspace{1.0cm}-\lambda^2_{N-1}\Delta_{N-1} \Delta_{N+1}\left[\begin{array}{cc}
           N-1& N  \\ 1&2
         \end{array}
       \right],\\
&&R_4=\lambda^2_{N}\Delta_{N}\left[\begin{array}{cc}
           N-1  \\ N
         \end{array}
       \right]\Delta_{N+1}\left[\begin{array}{cc}
           N&N+1  \\ 1&2
         \end{array}
       \right]\\
&&\hspace{1.0cm}-\lambda^2_{N-1}\Delta_{N-1} \Delta_{N+1}\left[\begin{array}{cc}
           N-1 &N+1 \\ 1 &2
         \end{array}
       \right].
 \end{eqnarray*}
Using the Jacobi identity, we have (proven in appendix)
  \begin{equation}
\begin{array}{lll}\label{jac2}
&&\hspace{-1.08cm}R_3=\frac{1}{\lambda_1^2\lambda_2^2\cdots\lambda_{N-2}^2}\Delta_{N+2}\left[\begin{array}{cc}
           N &N+2 \\ 1&2
         \end{array}
       \right]\Delta_{N+1}\left[\begin{array}{ccc}
           N-1 &N &N+1\\ 1&2&3
         \end{array}
       \right],\\
&&\hspace{-1.08cm}R_4=\frac{1}{\lambda_1^2\lambda_2^2\cdots\lambda_{N-2}^2}\Delta_{N+2}\left[\begin{array}{cc}
           N+1 &N+2 \\ 1&2
         \end{array}
       \right]\Delta_{N+1}\left[\begin{array}{ccc}
           N-1 &N &N+1 \\ 1&2&3
         \end{array}
       \right].
\end{array}
\end{equation}
Substituting (\ref{jac2}) into (\ref{g3g4}) and noting the first two equalities in (\ref{plu2}), we get
\begin{equation}
\begin{array}{lll}\label{sin7}
\sigma^{N+1}_{\widehat{N}}=\frac{\lambda_N(A_{N-1}B_N-A_NB_{N-1})\Delta_{N+2}\left[\begin{array}{cc}
           N &N+2 \\ 1&2
         \end{array}
       \right]}{\lambda_{N+1}(A_{N-1}D_N-B_{N-1}C_N)\Delta_{N+2}\left[\begin{array}{cc}
           N+1 &N+2 \\ 1&2
         \end{array}
       \right]}\\
      \hspace{1.0cm} =
\frac{\lambda_N}{\lambda_{N+1}}\frac{(A_{N-1}B_N-A_NB_{N-1})(C_{N}B_{N+1}-A_{N+1}D_{N})}
{(A_{N-1}D_{N}-B_{N-1}C_{N})(A_{N}B_{N+1}-B_{N}A_{N+1})},
\end{array}
\end{equation}
which proves (\ref{sin3}).

Furthermore, it follows from  inductive hypotheses, (\ref{sin2}), (\ref{jac}) and (\ref{plu2}) that
\begin{eqnarray*}\label{wn4}
&&\hspace{-0.8cm}w_{\widehat{N+1}}=\frac{\lambda_{N+1}\sigma^{N+1}_{\widehat{N}}w_{\widehat{N}}(w_{\widehat{N-1}N+1}-w_{\widehat{N-1}})
   +\lambda_{N}w_{\widehat{N-1}N+1}(w_{\widehat{N-1}}-w_{\widehat{N}})}
   {\lambda_{N+1}\sigma^{N+1}_{\widehat{N}}(w_{\widehat{N-1}N+1}-w_{\widehat{N-1}})
   +\lambda_{N}(w_{\widehat{N-1}}-w_{\widehat{N}})} \\
&&\hspace{0.2cm}=\frac{\frac{(A_{N-1}B_N-A_NB_{N-1})(C_{N}B_{N+1}-A_{N+1}D_{N})}
{(A_{N-1}D_{N}-B_{N-1}C_{N})(A_{N}B_{N+1}-B_{N}A_{N+1})}\frac{A_N}{B_N}(\frac{C_N}{D_N}-\frac{A_{N-1}}{B_{N-1}})
+\frac{C_N}{D_N}(\frac{A_{N-1}}{B_{N-1}}-\frac{A_{N}}{B_{N}})}{\frac{(A_{N-1}B_N-A_NB_{N-1})(C_{N}B_{N+1}-A_{N+1}D_{N})}
{(A_{N-1}D_{N}-B_{N-1}C_{N})(A_{N}B_{N+1}-B_{N}A_{N+1})}(\frac{C_N}{D_N}-\frac{A_{N-1}}{B_{N-1}})
+\frac{A_{N-1}}{B_{N-1}}-\frac{A_{N}}{B_{N}}} \\
&&\hspace{0.2cm}=\frac{\Delta_{N+2}\left[\begin{array}{cc}
           N+1  &N+2\\
           1 & 2\\
         \end{array}
       \right]\Delta_{N+1}\left[\begin{array}{cc}
           N  \\ 1
         \end{array}
       \right]-\Delta_{N+2}\left[\begin{array}{cc}
           N  &N+2\\
           1 & 2\\
         \end{array}
       \right]\Delta_{N+1}\left[\begin{array}{cc}
           N+1  \\ 1
         \end{array}
       \right]}{\Delta_{N+2}\left[\begin{array}{cc}
           N+1  &N+2\\
           1 & 2\\
         \end{array}
       \right]\Delta_{N+1}\left[\begin{array}{cc}
           N  \\ 2
         \end{array}
       \right]-\Delta_{N+2}\left[\begin{array}{cc}
           N  &N+2\\
           1 & 2\\
         \end{array}
       \right]\Delta_{N+1}\left[\begin{array}{cc}
           N+1  \\ 2
         \end{array}
       \right]}\\
&&\hspace{0.2cm}=\frac{A_{N+1}\left[\begin{array}{c}
           N+1  \\
           1 \\
         \end{array}
       \right]A_{N+1}\left[\begin{array}{c}
           N  \\ N+1
         \end{array}
       \right]-A_{N+1}\left[\begin{array}{c}
           N  \\
           1 \\
         \end{array}
       \right]A_{N+1}\left[\begin{array}{c}
           N+1  \\ N+1
         \end{array}
       \right]}
       {B_{N+1}\left[\begin{array}{c}
           N+1  \\
           1\\
         \end{array}
       \right]B_{N+1}\left[\begin{array}{c}
           N  \\ N+1
         \end{array}
       \right]-B_{N+1}\left[\begin{array}{c}
           N  \\
           1\\
         \end{array}
       \right]B_{N+1}\left[\begin{array}{c}
           N+1  \\ N+1
         \end{array}
       \right]}\\
&&\hspace{0.2cm}\xlongequal[identity]{Jacobi }\frac{A_{N+1}A_{N+1}\left[\begin{array}{c}
           N \\
           1\\
         \end{array}
       \right]}{B_{N+1}B_{N+1}\left[\begin{array}{cc}
           N  \\
           1 \\
         \end{array}
       \right]} \\
&&\hspace{0.2cm}=\frac{A_{N+1}}{B_{N+1}},
\end{eqnarray*}
which completes the proof of Theorem 1.

Now, according to Theorem 1, it directly infers from (\ref{an2}) that
\begin{equation}\label{an4}
a_{\widehat{N}}=\frac{A_{N-1}A_N-B_{N-1}B_N}{A_{N-1}B_N-B_{N-1}A_N}.
\end{equation}
Thus, we may summarize what we have obtained as the following Theorem.

\begin{thm}The GX system admits the $N$-BT
 \begin{eqnarray}
 \begin{aligned}
 &x_{[\widehat{N}]}=x-\frac{1}{2}{\rm ln} |\frac{A_N+B_N}{A_N-B_N}|,\quad \quad \quad  t_{[\widehat{N}]}=t, \\
 &u_{[\widehat{N}]}=\frac{1}{c_{\widehat{N}}}(u_{[\widehat{N-1}]}a_{\widehat{N}}-\frac{u_{[\widehat{N-1}],x}}{x_{[\widehat{N-1}],x}}), \\ &v_{[\widehat{N}]}=\frac{c_{\widehat{N}}}{a^2_{\widehat{N}}-1}[v_{[\widehat{N-1}]}a_{\widehat{N}}-\frac{v_{[\widehat{N-1}],x}}{x_{[\widehat{N-1}],x}}
 -\frac{1}{\lambda^2_Nm_{[\widehat{N-1}]}}(\frac{a_{\widehat{N},x}}{x_{[\widehat{N-1}],x}}+a^2_{\widehat{N}}-1)],
 \end{aligned}
\end{eqnarray}
where $a_{\widehat{N}}$ is given by \eqref{an4}, $c_{\widehat{N}}=\sqrt{|a_{\widehat{N}}^2-1|},$ and $m_{\widehat{N-1}}=u_{\widehat{N-1}}-\frac{1}{x_{\widehat{[N-1]},x}}(\frac{u_{[\widehat{N-1}],x}}{x_{[\widehat{N-1}],x}})_x $.
\end{thm}

\section{Exact solutions}
As an application of the BT, we shall deduce some exact solutions of the GX system. Choose $u=u_0, v=v_0, u_0v_0\neq0$ as an initial solution of the GX system. Let $\alpha_j, \beta_j, -\alpha_j-\beta_j, (\ 1\leq j\leq N) $ be three roots of the equation $\gamma^3-\gamma-\lambda_j^2u_0v_0=0,$ and $f_j$ be solutions of the following system \begin{eqnarray}
\begin{aligned}
&\varphi_{xxx}-\varphi_x-\lambda_j^2u_0v_0\varphi=0,\\ &\varphi_t-\frac{1}{\lambda_j^2}\varphi_{xx}+u_0v_0\varphi_x+\frac{1}{\lambda_j^2}\varphi=0,
\end{aligned}
\end{eqnarray}
which is a scalar form of \eqref{gxlax} at $\lambda=\lambda_j, u=u_0,v=v_0$.

{\bf Example 1:} 1-soliton solutions.

 If $27\lambda_1^4u_0^2v_0^2-4<0$, then $\alpha_1, \beta_1, -\alpha_1-\beta_1$ are three different real roots.  We take
\begin{equation*}
f_1=e^{\frac{\xi_1+\eta_1}{2}}(e^{\theta_1}+\delta_1e^{-\theta_1}),
\end{equation*}
where $\xi_1=\alpha_1 x-\frac{\lambda_1^2 u_0^2v_0^2}{\alpha_1^2}t+\xi_{10},\
\eta_1=\beta_1 x-\frac{\lambda_1^2 u_0^2v_0^2}{\beta_1^2}t+\eta_{10},\ \theta_1=\frac{\mu_1}{2}[x+\frac{u_0v_0(4-\mu_1^2)}{\mu_1^2-1}t]+\theta_{10}, \  \mu_1=\alpha_1-\beta_1,\ \delta_1=\pm1,\ \theta_{10}=\frac{1}{2}(\xi_{10}-\eta_{10})$, and $\xi_{10},\eta_{10}$ are two constants. For convenience, assume $\mu_1>0$ and let $\nu_1=\alpha_1+\beta_1$, it infers $\mu_1=\sqrt{4-3\nu_1^2}$.
Then
\begin{equation*}
a_1=\frac{(\mu_1+\nu_1)e^{\theta_1}+\delta_1(\nu_1-\mu_1)e^{-\theta_1}}{2(e^{\theta_1}+\delta_1e^{-\theta_1})}
=\left\{\begin{array}{lll}
\frac{\nu_1+\mu_1\tanh\theta_1}{2},\ \ \delta_1=1,\\
\frac{\nu_1+\mu_1\coth\theta_1}{2},\ \ \delta_1=-1,
\end{array}
\right.
\end{equation*}
which together with Proposition 2 yields tanh type 1-soliton solution
\begin{eqnarray}
\begin{aligned}\label{s1}
&x_{[1]}=x-\frac{1}{2}\ln|\frac{\nu_1+2+\mu_1\tanh\theta_1}{\nu_1-2+\mu_1\tanh\theta_1}|,\\
&u_{[1]}=\frac{u_0(\nu_1+\mu_1\tanh\theta_1)}{\sqrt{|(\nu_1+\mu_1\tanh\theta_1)^2-4|}},\\
&v_{[1]}=-\frac{v_0\nu_1(\nu_1^2-2+\mu_1\mu_1\tanh\theta_1)\sqrt{|(\nu_1+\mu_1\tanh\theta_1)^2-4|}}
{(1-\nu_1^2)[(\nu_1+\mu_1\tanh\theta_1)^2-4]},
\end{aligned}
\end{eqnarray}
and coth type 1-soliton solution
\begin{eqnarray}
\begin{aligned}\label{s2}
&x_{[1]}=x-\frac{1}{2}\ln|\frac{\nu_1+2+\mu_1\coth\theta_1}{\nu_1-2+\mu_1\coth\theta_1}|,\\
&u_{[1]}=\frac{u_0(\nu_1+\mu_1\coth\theta_1)}{\sqrt{|(\nu_1+\mu_1\coth\theta_1)^2-4|}},\\
&v_{[1]}=-\frac{v_0\nu_1(\nu_1^2-2+\mu_1\nu_1\coth\theta_1)\sqrt{|(\nu_1+\mu_1\coth\theta_1)^2-4|}}
{(1-\nu_1^2)[(\nu_1+\mu_1\coth\theta_1)^2-4]}.
\end{aligned}
\end{eqnarray}
It is easy to see that $x_{[1]} \to \pm \infty$ when $x\to \pm \infty$ in \eqref{s1} and \eqref{s2}. Further analysis shows the map from $x_{[1]}$ to $x$ in \eqref{s1} is bijective and $x_{[1]}, u_{[1]}, v_{[1]}$ are nonsingular when $1<|\nu_1|<\frac{2}{\sqrt{3}}$, while in \eqref{s2} $x_{[1]}, u_{[1]}, v_{[1]}$ are singular for any $\nu_1$. It infers that \eqref{s1} gives smooth soliton solutions for $1<|\nu_1|<\frac{2}{\sqrt{3}}$ and \eqref{s2} gives singular solutions to which we do not pay more attention. The profiles of the 1-soliton solutions \eqref{s1} are shown in Fig. 2-3.

\begin{figure}[htb]
\begin{minipage}[b]{.5\textwidth}
 \centering
\includegraphics[scale=0.45]{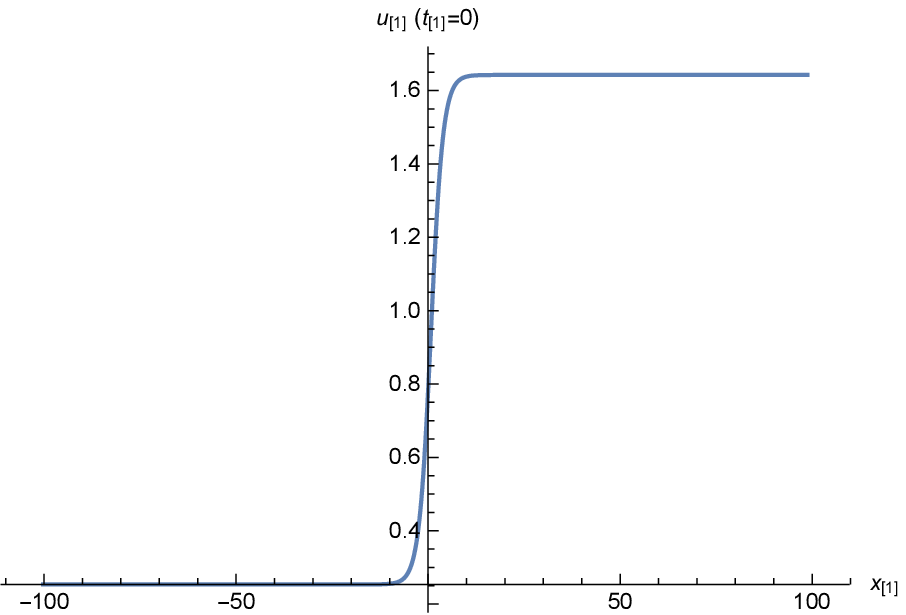}
  \end{minipage} \quad
\begin{minipage}[b]{.5\textwidth}
 \includegraphics[scale=0.45]{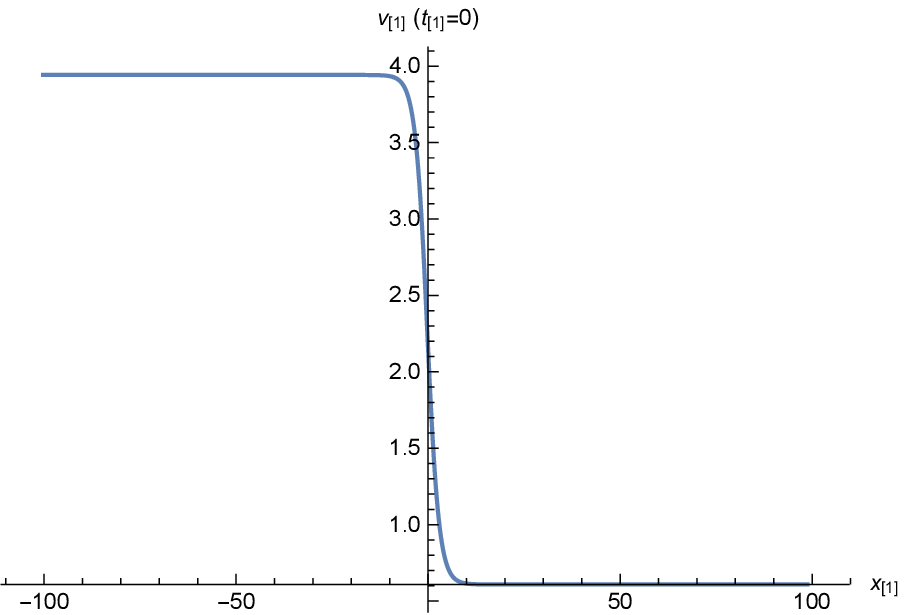}
 \end{minipage}
\caption{The profiles of smooth 1-soliton solution \eqref{s1} at $u_0=1, v_0=1, \nu_1=1.1. $}
\label{p1}
\end{figure}

\begin{figure}[htb]
\begin{minipage}[b]{.5\textwidth}
 \centering
\includegraphics[scale=0.45]{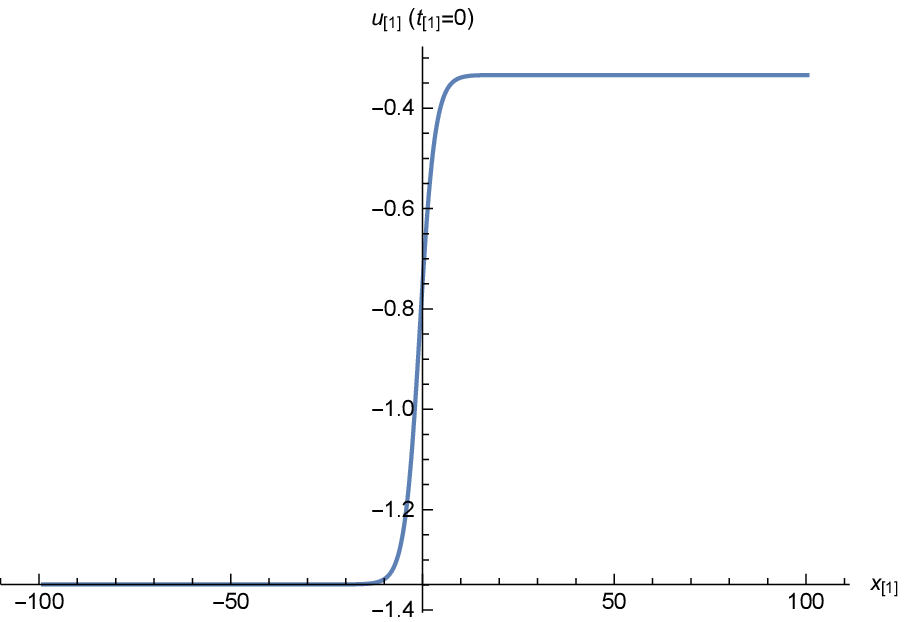}
  \end{minipage} \quad
\begin{minipage}[b]{.5\textwidth}
 \includegraphics[scale=0.45]{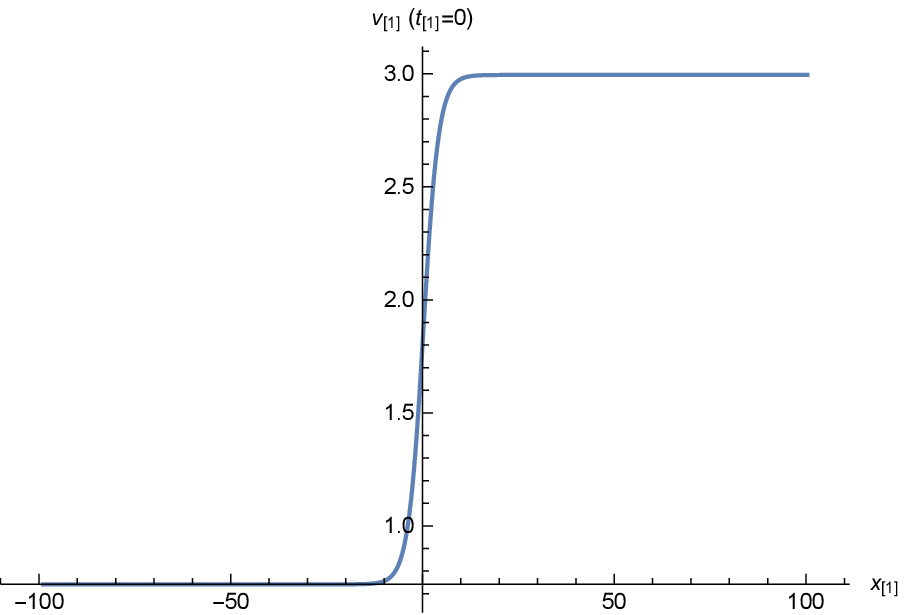}
 \end{minipage}
\caption{The profiles of smooth 1-soliton solution \eqref{s1} at $u_0=1, v_0=-1, \nu_1=-1.12. $}
\label{p2}
\end{figure}

{\bf Example 2:} 2-soliton solutions and their interactions.

Applying the Proposition 3, we have 2-soliton solution
\begin{eqnarray}
\begin{aligned}\label{s11}
 &x_{[12]}=x-\frac{1}{2}{\rm ln} \lvert \frac{(a_1+1)(a_{12}+1)}{(a_1-1)(a_{12}-1)}\rvert,\quad \quad \quad \quad \quad  t_{[12]}=t, \\
 &u_{[12]}=\frac{1}{c_1c_{12}}[u_0(a_1a_{12}+1)-\frac{a_1^2-1}{\lambda_1b_1}),\\
 &v_{[12]}=\frac{c_1c_{12}}{(a_1^2-1)(a_{12}^2-1)}[v_0(a_1a_{12}+1)-\frac{b_1}{\lambda_1}(a_1+a_{12})]\\
& \quad\quad\quad -\frac{c_1c_{12}}{(a_2-a_1)(a_{12}^2-1)}(\frac{b_1}{\lambda_1}-\frac{b_2}{\lambda_2})
\end{aligned}
\end{eqnarray}
with $c_1=\sqrt{|a_1^2-1|},\ c_{12}=\sqrt{|a_{12}^2-1|}.$
We call \eqref{s11} tanh-tanh type and tanh-coth type 2-soliton solution respectively when
\begin{eqnarray*}
\begin{aligned}
&a_1=\frac{\nu_1+\mu_1\tanh\theta_1}{2}, \quad \quad  \quad  \quad  \quad  \ a_2=\frac{\nu_2+\mu_2\tanh\theta_1}{2},\\
&b_1=\frac{-\nu_1(\nu_1-\mu_1\tanh\theta_1)}{2\lambda_1u_0}, \quad \quad  \quad b_2=\frac{-\nu_2(\nu_2-\mu_2\tanh\theta_2)}{2\lambda_2u_0},\\
&a_{12}=\frac{4-(\nu_1+\mu_1\tanh\theta_1)(\nu_2+\mu_2\tanh\theta_2)}{2(\nu_2-\nu_1+\mu_2\tanh\theta_2-\mu_1\tanh\theta_1)}\\
&\quad\quad \ -\frac{\nu_2(\nu_2-\mu_2\tanh\theta_2)[4-(\nu_1+\mu_1\tanh\theta_1)^2]}{2\nu_1(\nu_1-\mu_1\tanh\theta_1)(\nu_2-\nu_1+\mu_2\tanh\theta_2-\mu_1\tanh\theta_1)},
\end{aligned}
\end{eqnarray*}
and
\begin{eqnarray*}
\begin{aligned}
&a_1=\frac{\nu_1+\mu_1\tanh\theta_1}{2},  \quad \quad  \quad \quad  \quad \ a_2=\frac{\nu_2+\mu_2\coth\theta_1}{2},\\
&b_1=\frac{-\nu_1(\nu_1-\mu_1\tanh\theta_1)}{2\lambda_1u_0}, \quad \quad  \quad b_2=\frac{-\nu_2(\nu_2-\mu_2\coth\theta_2)}{2\lambda_2u_0},\\
&a_{12}=\frac{4-(\nu_1+\mu_1\tanh\theta_1)(\nu_2+\mu_2\coth\theta_2)}{2(\nu_2-\nu_1+\mu_2\coth\theta_2-\mu_1\tanh\theta_1)}\\
&\quad\quad \ -\frac{\nu_2(\nu_2-\mu_2\coth\theta_2)[4-(\nu_1+\mu_1\tanh\theta_1)^2]}{2\nu_1(\nu_1-\mu_1\tanh\theta_1)(\nu_2-\nu_1+\mu_2\coth\theta_2-\mu_1\tanh\theta_1)}.
\end{aligned}
\end{eqnarray*}
Here $ \theta_1=\frac{\mu_1}{2}[x+\frac{u_0v_0(4-\mu_1^2)}{\mu_1^2-1}t]+\theta_{10}, \ \theta_2=\frac{\mu_2}{2}[x+\frac{u_0v_0(4-\mu_2^2)}{\mu_2^2-1}t]+\theta_{20}, \ \mu_1=\sqrt{4-3\nu_1^2}, \ \mu_2=\sqrt{4-3\nu_2^2},
\ \lambda_1^2=\frac{\nu_1(1-\nu_1^2)}{u_0v_0},\  \lambda_2^2=\frac{\nu_2(1-\nu_2^2)}{u_0v_0},$ and $\nu_1, \  \nu_2$ are two constants.

Analysis shows that  tanh-tanh type 2-soliton solution gives smooth kink-antikink or antikink-kink solution if $1<|\nu_1|,|\nu_2| <\frac{2}{\sqrt{3}},\  \nu_1\nu_2<0$. Especially and interestingly, one finds that the tanh-tanh type 2-soliton solution becomes bell-shaped 1-soliton solutions when $\nu_1=-\nu_2$.  Moreover, tanh-coth type 2-soliton solution gives kink-kink or antikink-antikink solutions when $ 1<|\nu_2|<|\nu_1|<\frac{2}{\sqrt{3}},\ \nu_1\nu_2>0$.
 The profiles of the 2-soliton solutions \eqref{s11} and their interactions are shown in Fig. 4-6.
 \begin{figure}[htb]
\begin{minipage}[b]{.3\textwidth}
 \centering
\includegraphics[scale=0.45]{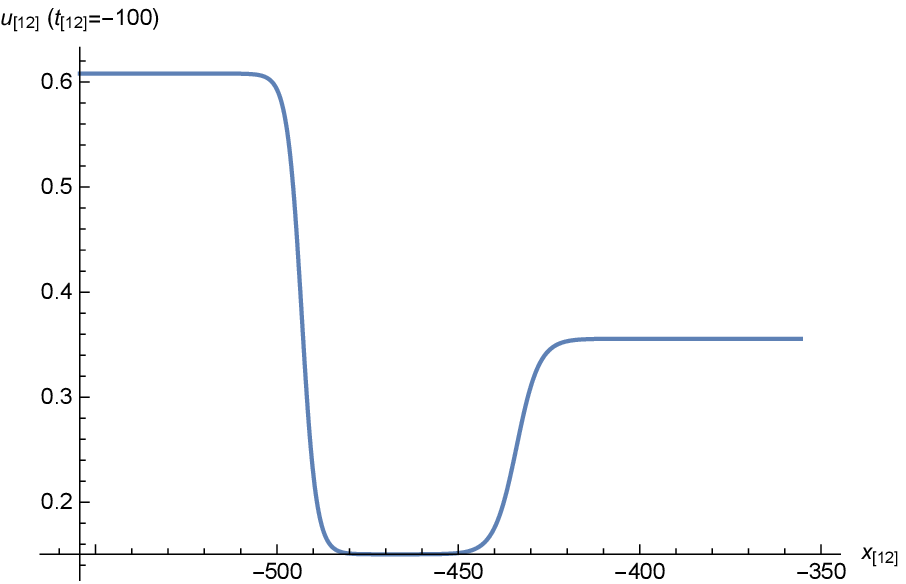}
  \end{minipage} \quad
\begin{minipage}[b]{.3\textwidth}
 \includegraphics[scale=0.45]{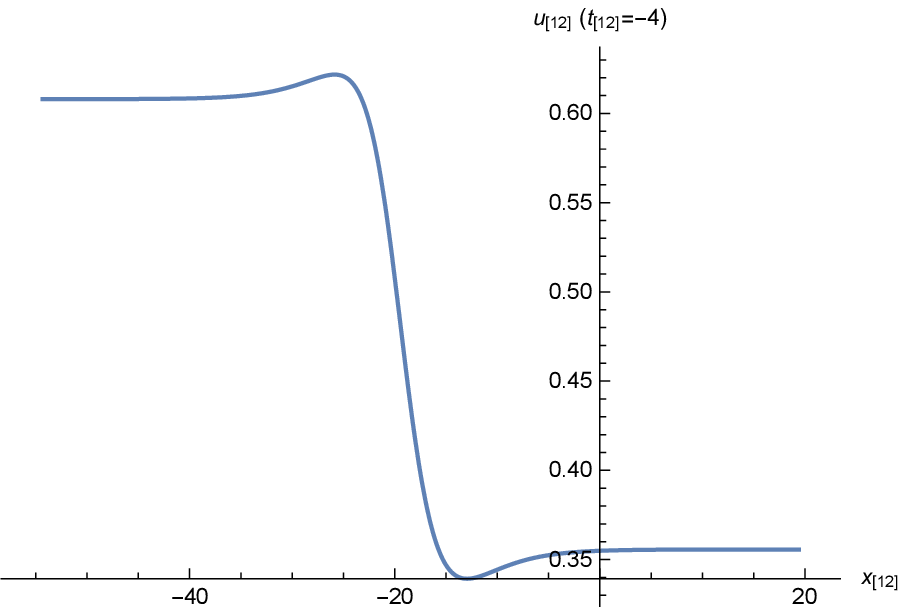}
 \end{minipage}\quad
 \begin{minipage}[b]{.3\textwidth}
 \includegraphics[scale=0.45]{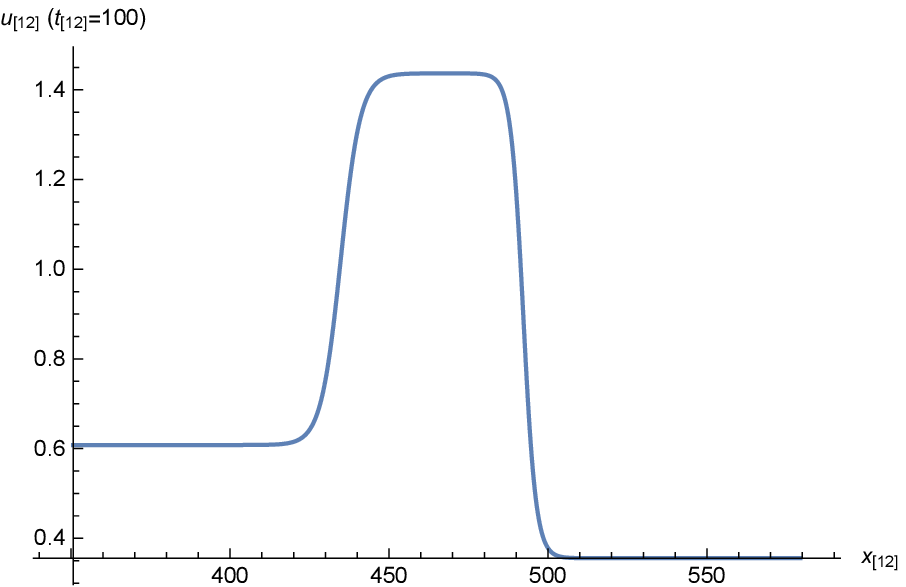}
 \end{minipage}\\
 \begin{minipage}[b]{.3\textwidth}
 \centering
\includegraphics[scale=0.45]{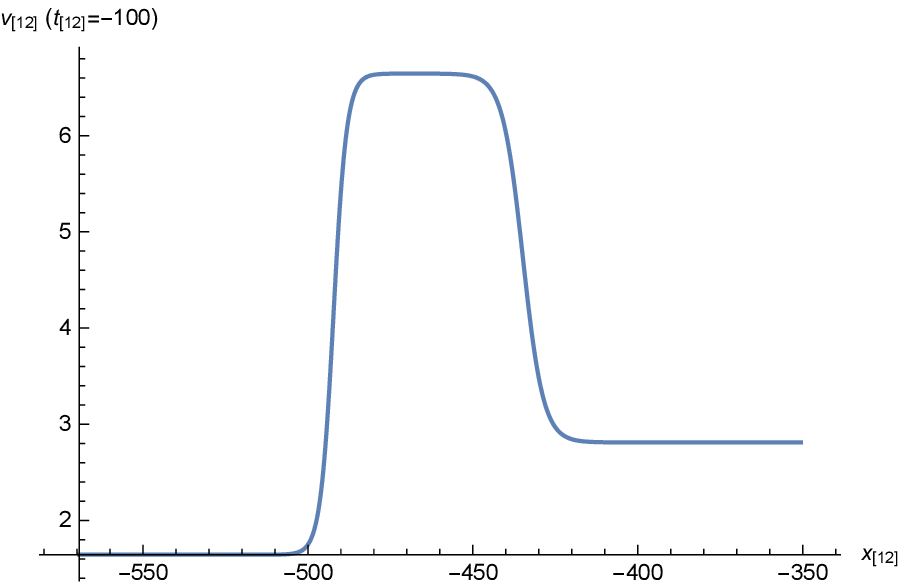}
  \end{minipage} \quad
\begin{minipage}[b]{.3\textwidth}
 \includegraphics[scale=0.45]{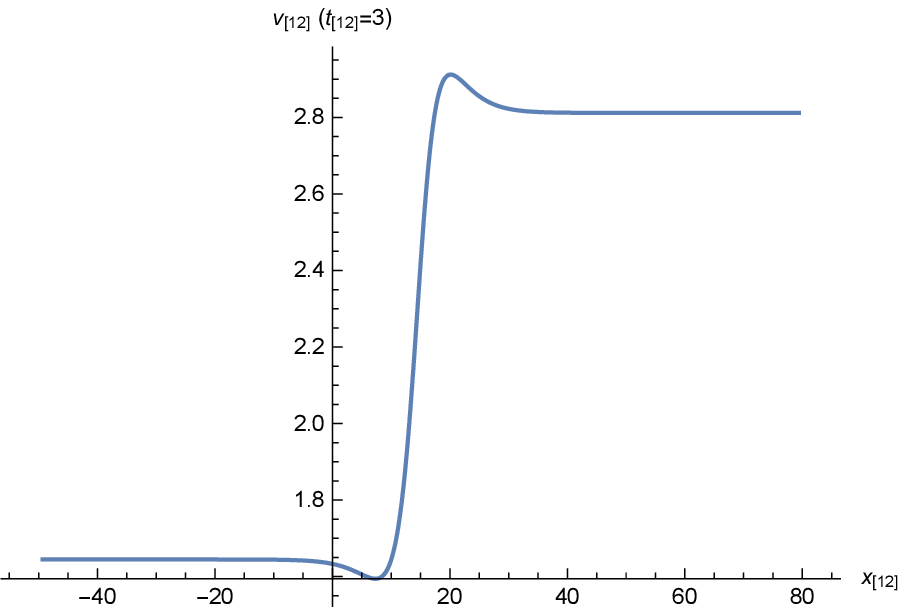}
 \end{minipage}\quad
 \begin{minipage}[b]{.3\textwidth}
 \includegraphics[scale=0.45]{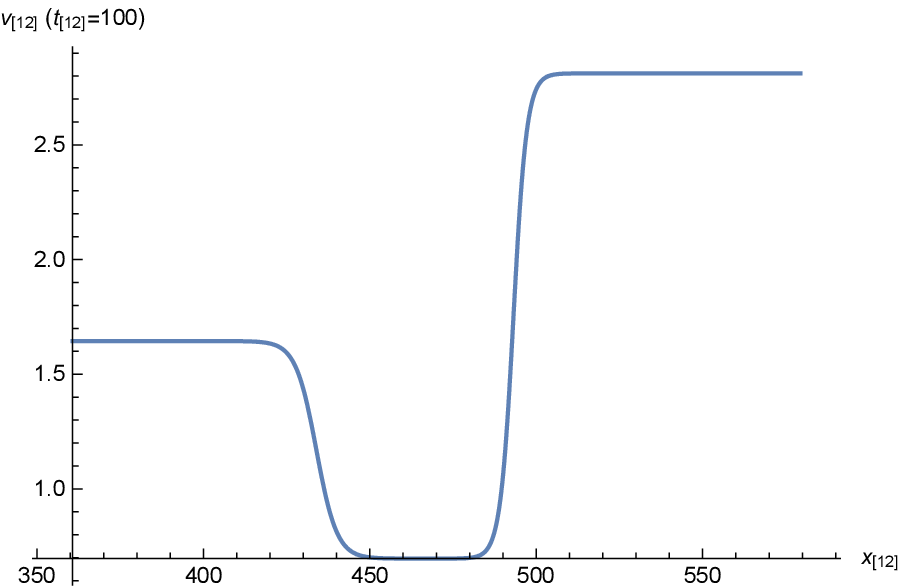}
 \end{minipage}
\caption{The antikink-kink and kink-antikink solutions at $u_0=-1, v_0=-1,  \nu_1=-1.12, \nu_2=1.14, \theta_{10}=\theta_{20}=0. $}
\label{p3}
\end{figure}

\begin{figure}[htb]
\begin{minipage}[b]{.3\textwidth}
 \centering
\includegraphics[scale=0.45]{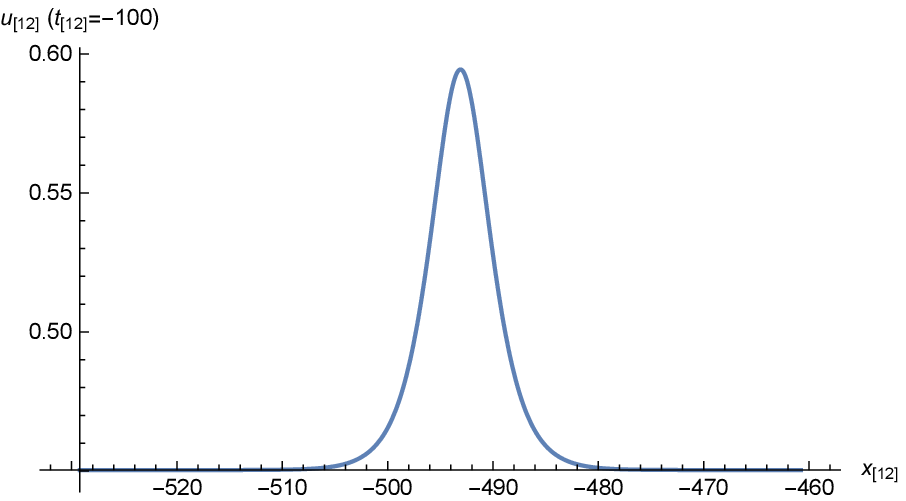}
  \end{minipage} \quad
\begin{minipage}[b]{.3\textwidth}
 \includegraphics[scale=0.45]{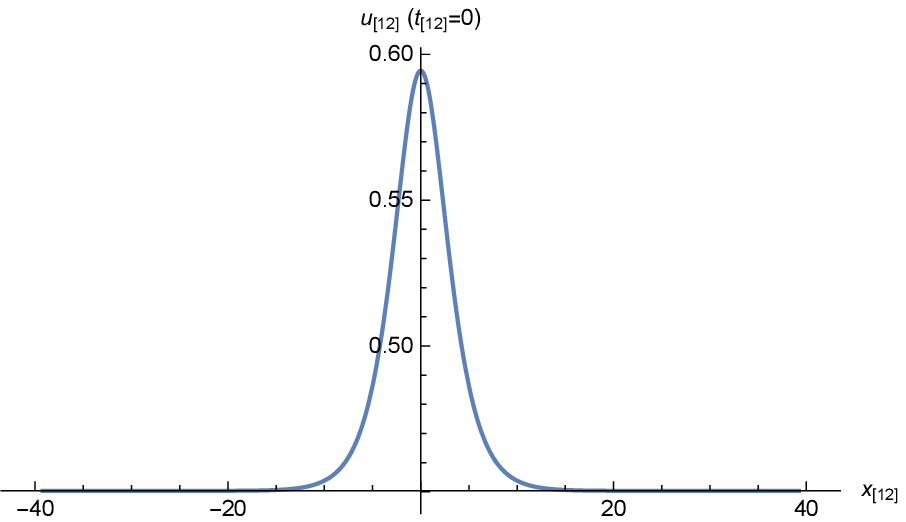}
 \end{minipage}\quad
 \begin{minipage}[b]{.3\textwidth}
 \includegraphics[scale=0.45]{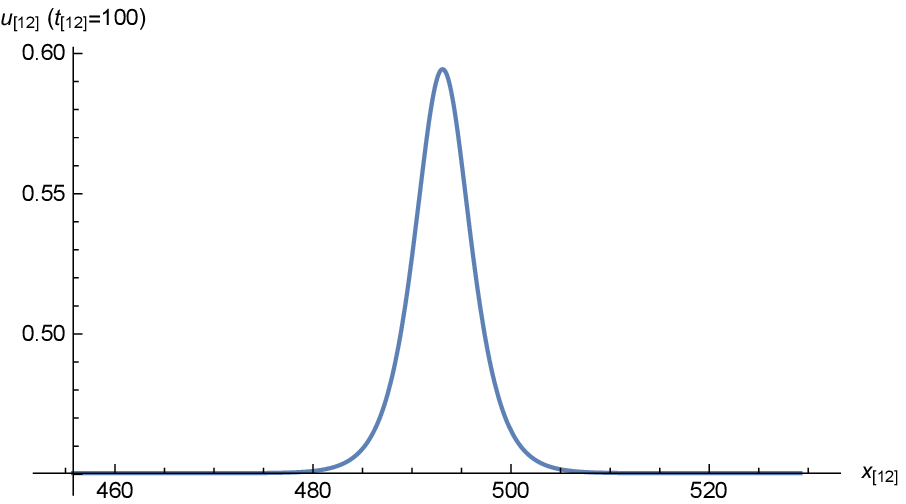}
 \end{minipage}\\
 \begin{minipage}[b]{.3\textwidth}
 \centering
\includegraphics[scale=0.45]{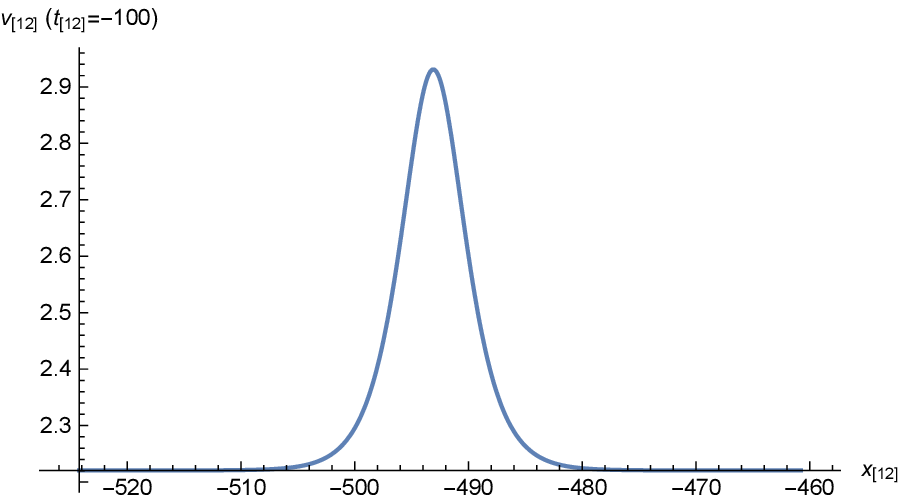}
  \end{minipage} \quad
\begin{minipage}[b]{.3\textwidth}
 \includegraphics[scale=0.45]{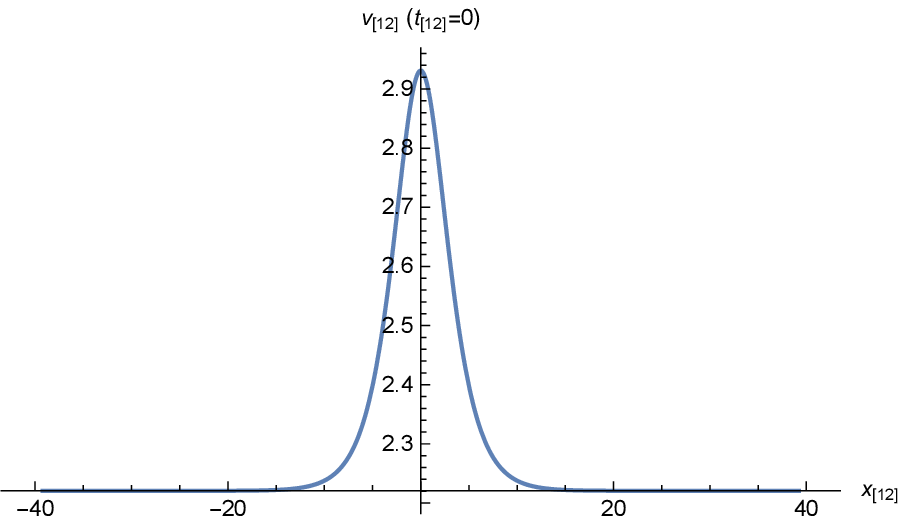}
 \end{minipage}\quad
 \begin{minipage}[b]{.3\textwidth}
 \includegraphics[scale=0.45]{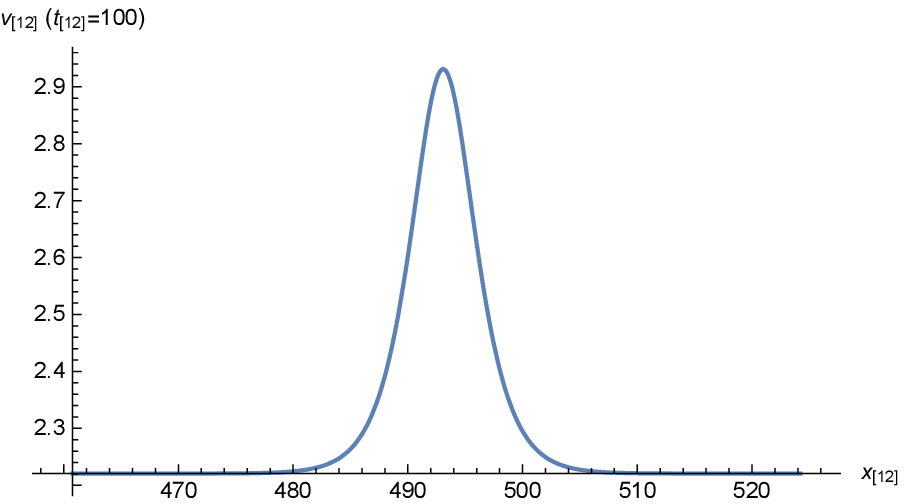}
 \end{minipage}
\caption{The bell-shaped 1-soliton solution at $u_0=-1, v_0=-1, \nu_1=-\nu_2=-1.12, \theta_{10}=\theta_{20}=0. $}
\label{p4}
\end{figure}

 \begin{figure}[htb]
 \begin{minipage}[b]{.3\textwidth}
 \centering
\includegraphics[scale=0.45]{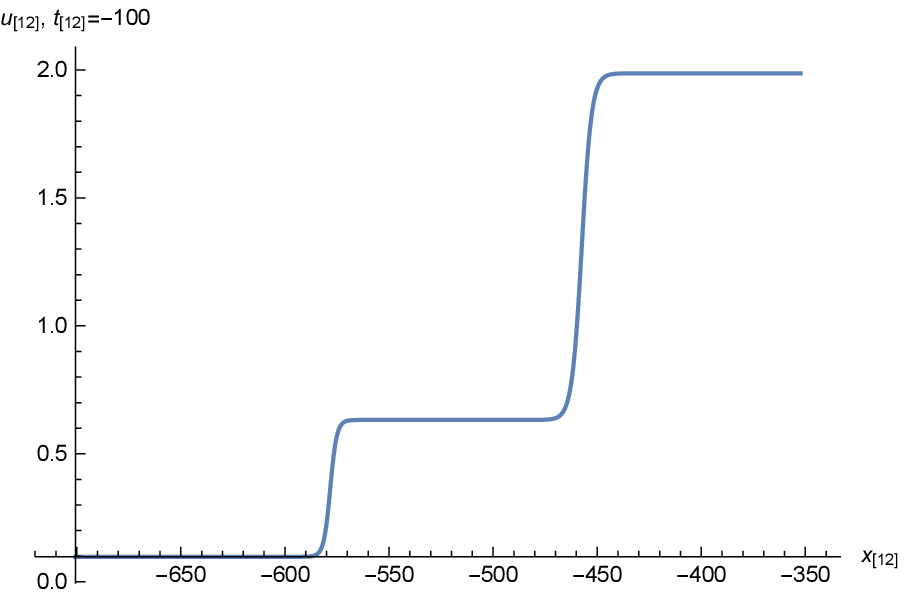}
  \end{minipage} \quad
\begin{minipage}[b]{.3\textwidth}
 \includegraphics[scale=0.45]{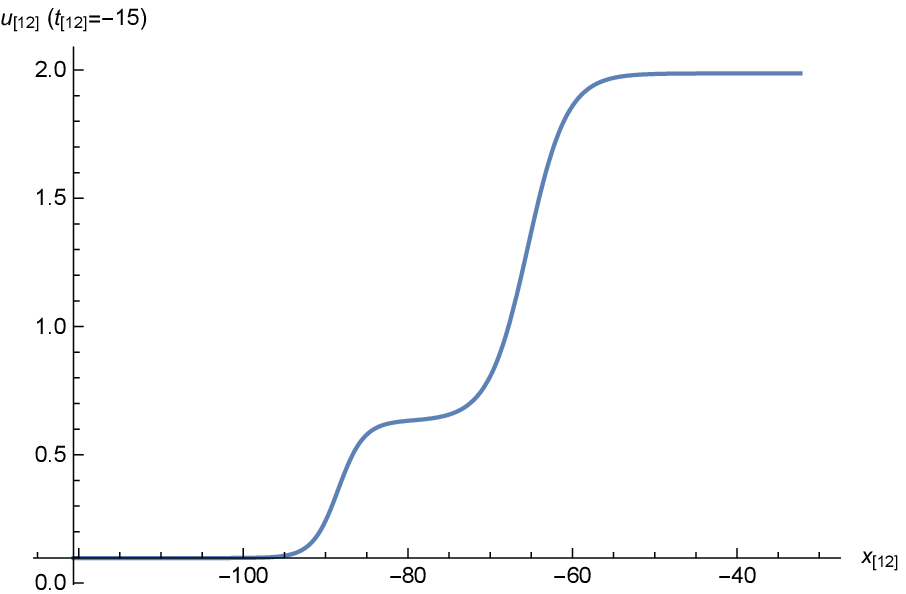}
 \end{minipage}\quad
 \begin{minipage}[b]{.3\textwidth}
 \includegraphics[scale=0.45]{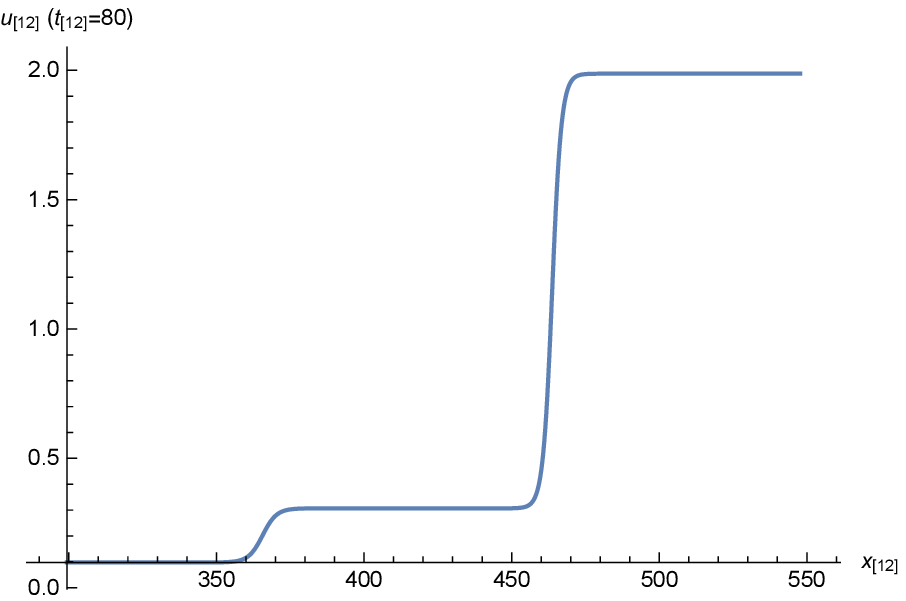}
 \end{minipage}\\
 \begin{minipage}[b]{.3\textwidth}
 \centering
\includegraphics[scale=0.45]{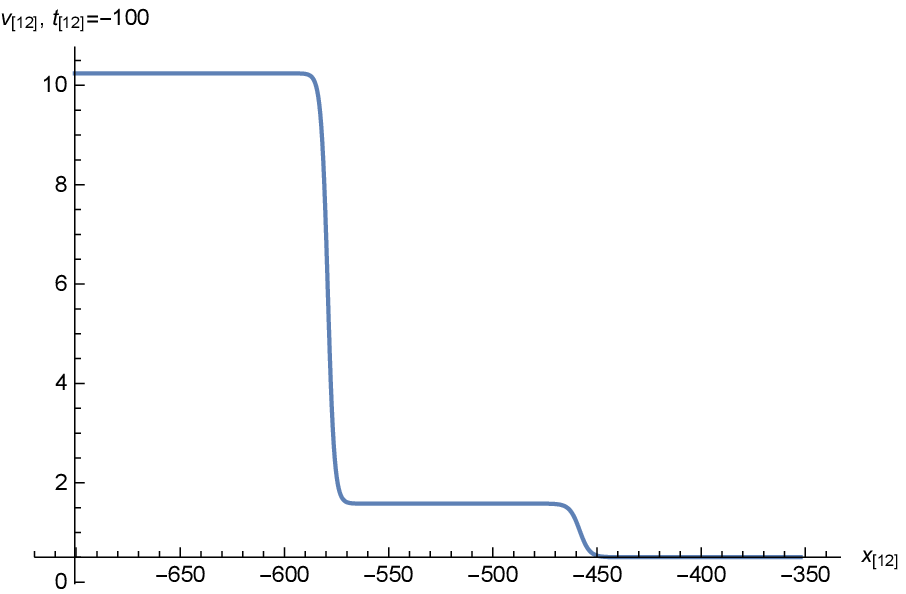}
  \end{minipage} \quad
\begin{minipage}[b]{.3\textwidth}
 \includegraphics[scale=0.45]{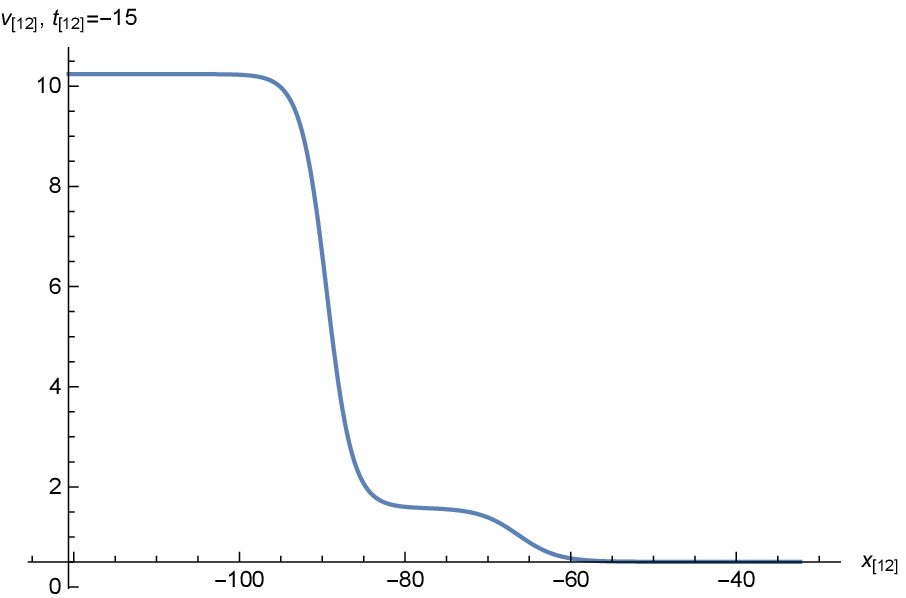}
 \end{minipage}\quad
 \begin{minipage}[b]{.3\textwidth}
 \includegraphics[scale=0.45]{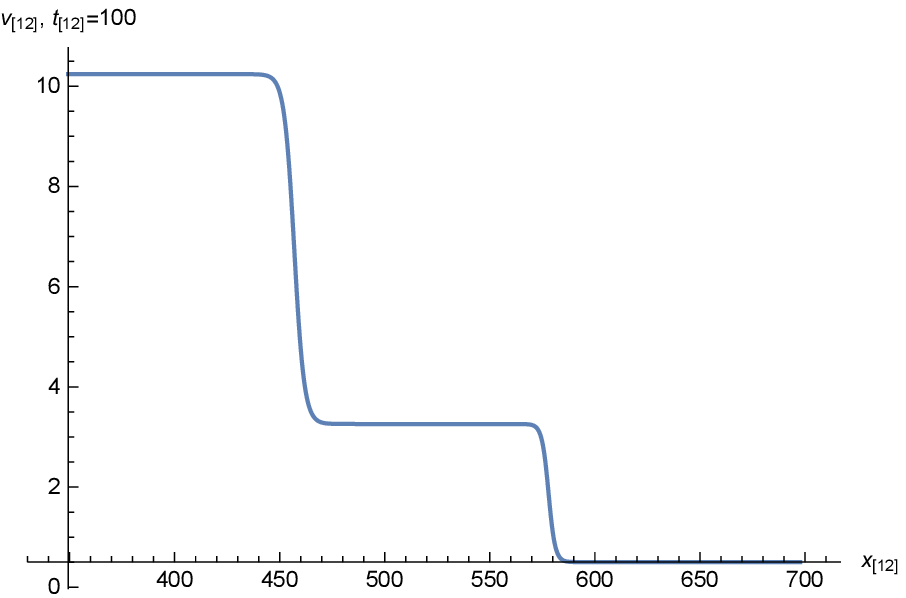}
 \end{minipage}
\caption{The kink-kink and antikink-antikink solutions at $u_0=1, v_0=1,\nu_1=1.13, \nu_2=1.1, \theta_{10}=\theta_{20}=0. $}
\label{p5}
\end{figure}

\section{B\"{a}cklund transformations for the Degasperis-Procesi and Novikov equations}
\subsection{The Degasperis-Procesi case}
As we know, the GX system reduces to the DP equation as $v=1$. Therefore,
we shall study BT for the DP equation. The DP equation \eqref{dp} possesses a Lax pair \cite{Deg2}
\begin{equation}\label{dplax}
\psi_x=U_1\psi,\quad \quad \quad \psi_t=V_1\psi,
\end{equation}where $\psi=(\psi_1,\psi_2,\psi_3)^{T}$ and
\begin{equation*}
U_1=\left[
\begin{matrix}
                   0 & \lambda m & 1\\
                   0 & 0 & \lambda \\
                   1 & 0 & 0 \\
                  \end{matrix}
                \right],\quad \quad \quad V_1=\left[
\begin{matrix}
                  -u_x & \frac{u_x}{\lambda}-\lambda um  & 0\\
                   \frac{1}{\lambda} & -\frac{1}{\lambda^2}+u_x & -\lambda u \\
                   -u &  \frac{u}{\lambda} & 0 \\
                  \end{matrix}
                \right].
\end{equation*}
Naturally, the reciprocal transformation (\ref{gxre}) of the Geng-Xue system  reduces to that of the DP equation
\begin{equation}\label{dpre}
dy=qdx-uqdt,\quad \quad \quad d\tau=dt.
\end{equation}
with $q=m^{\frac{1}{3}}$.
Applying this  transformation, the Lax presentation (\ref{dplax}) is converted  to
\begin{equation}\label{adplax}
\psi_y=F_1\psi,\quad \quad \quad \psi_\tau=G_1\psi,
\end{equation}where
\begin{equation*}
F_1=\left[
\begin{matrix}
                   0 & \lambda q^2  & \frac{1}{q}\\
                   0 & 0 & \frac{\lambda}{q} \\
                   \frac{1}{q} & 0 &0\\
                  \end{matrix}
                \right],\quad \quad \quad G_1=\left[
\begin{matrix}
                -u_yq & \frac{u_yq}{\lambda}  & u\\
                   \frac{1}{\lambda} & u_yq-\frac{1}{\lambda^2} & 0 \\
                  0 &   \frac{u}{\lambda} & 0 \\
                  \end{matrix}
                \right].
\end{equation*} The compatibility condition of Lax pair (\ref{adplax}) yields the associated DP (aDP) equation \cite{wang1,Rasin2}
\begin{equation}\label{aDP}
q_{\tau}=-u_yq^2,\quad \quad \quad u-q^3-q(u_yq)_y=0.
\end{equation}
It is straightforward to verify that, under a gauge transformation $\chi=q\psi_2$, the scalar form of spectral problem in (\ref{adplax}) is converted to
\begin{equation*}
(\partial_y^3+U_1\partial_y+\frac{1}{2}U_{1y})\chi=\lambda_1^2\chi,\quad \quad \quad U_1=-2\frac{q_{yy}}{q}+\frac{q_y^2-1}{q^2},
\end{equation*}
 which is just the classical spectral problem of the KK hierarchy \cite{Leble}.
Making use of the DT for the KK hierarchy, we get the following Proposition.
\begin{prop}
The Lax pair (\ref{adplax}) is covariant under the DT
\begin{equation}\label{adpdt1}
 \begin{array}{l}
 \psi_{[1]}=T\psi,\\
 T=I+\left[
      \begin{array}{ccc}
        \frac{(\lambda_1\sigma_1-\sigma_2)(\lambda_1^2-\sigma_2^2)}{\lambda_1^2-2\lambda_1\sigma_1\sigma_2+\sigma_2^2} & \sigma_2 & 0 \\
        0 & \lambda\sigma_2 & 0 \\
        0 & 0 & \lambda_1 \\
      \end{array}
    \right]\left[
\begin{matrix}
                 \lambda^2 & \lambda& -\lambda^2 \\
               \lambda_1^2 & -\lambda&-\lambda_1^2\\
                   \lambda_1^2 &  -\lambda & -\lambda_1^2 \\
                  \end{matrix} \right]T_1,\\
q_{[1]}=\frac{q(\lambda_1^2-\sigma_2^2)}{\lambda_1^2-2\lambda_1\sigma_1\sigma_2+\sigma_2^2},\\
u_{[1]}=u+\frac{2(\lambda_1^2u_yh\sigma_2-\lambda_1^3u+\lambda_1\sigma_1-\sigma_2)}{\lambda_1(\lambda_1^2-\sigma_2^2)},
\end{array}
\end{equation}
 or the DT
\begin{eqnarray}\label{adpdt2}
 \begin{array}{l}\psi_{[1]}=\tilde{T}\psi,\ \ \ \ \tilde{T}={\rm diag}[-1,-1,1]T,\\
q_{[1]}=-\frac{q(\lambda_1^2-\sigma_2^2)}{\lambda_1^2-2\lambda_1\sigma_1\sigma_2+\sigma_2^2},\\
u_{[1]}=-u+\frac{2(\sigma_2-\lambda_1\sigma_1+\lambda^3_1u-\lambda_1^2u_yh\sigma_2)}{\lambda_1(\lambda_1^2-\sigma_2^2)},\end{array}
\end{eqnarray}
where $T_1=\frac{2
 }{(\lambda^2+\lambda_1^2)(\lambda_1^2-\sigma_2^2)}{\rm diag}[\sigma_2,\lambda_1\sigma_1-\sigma_2,\lambda_1]$, $\sigma_i=\frac{g_i}{g_3},i=1,2$, and $(g_1,g_2,g_3)^{T}$ is a special solution of the linear system (\ref{adplax}) at $\lambda=\lambda_1$.
\end{prop}

Since the process is very similar, we just consider the second DT in Proposition 4. It is easy to check that
\begin{equation}\label{dptheta}
\frac{1}{q_{[1]}}=\frac{1}{q}+\frac{2\vartheta_y}{\vartheta^2-1},\quad \quad \quad u_{[1]}=u+\frac{2\vartheta_\tau}{\vartheta^2-1},\quad \quad \quad  \vartheta=\frac{\lambda_1}{\sigma_2}.
\end{equation}Then with the aid of (\ref{dpre}), we obtain
$$dx_{[1]}=d(x-{\rm ln} \lvert \frac{\vartheta+1}{\vartheta-1}\rvert),$$
which infers that
\begin{equation*}
x_{[1]}=x-{\rm ln} \lvert \frac{\vartheta+1}{\vartheta-1}\rvert.
\end{equation*}
Here the integration constant is taken to be zero.
\begin{cor}
The DP equation has a BT of the form
\begin{eqnarray}
 \begin{aligned}
 &x_{[1]}=x-{\rm ln} \lvert \frac{\vartheta+1}{\vartheta-1}\rvert,\quad\quad\quad t_{[1]}=t, \\
&u_{[1]}=u-\frac{2\vartheta}{\lambda_1^2}+\frac{2\lambda_1^2(u-u_x\vartheta)-2\vartheta\vartheta_x}{\lambda^2_1(\vartheta^2-1)}, \end{aligned}
\end{eqnarray}where $\vartheta$ is determined by
\begin{equation*}
\vartheta_{xx}=\lambda_1^2m-3\vartheta\vartheta_x+\vartheta-\vartheta^3, \quad \quad
\vartheta_t=u-(u\vartheta)_x+\lambda_1^{-2}(\vartheta-\vartheta^3-\vartheta\vartheta_x).
\end{equation*}
\end{cor}

{\bf Remark:} In fact, considering the first DT in Proposition 4, one may get an equivalent BT to the one in \cite{mw}, which are related by $a=\frac{f_2^2p^2}{\int f_2^2p^2dy}$. Moreover, one can also discuss the $N$-BT for the DP equation like Section 2 which
will not reproduce here.

\subsection{The Novikov equation}
Now we consider BT for the Novikov equation (\ref{novikov}), which is another reduction of the Geng-Xue system as $u=v$. The Novikov equation admits the following Lax pair \cite{Hone}
\begin{equation}\label{novlax}
\psi_x=U_2\psi,\quad \quad \quad \psi_t=V_2\psi,
\end{equation}where $\psi=(\psi_1,\psi_2,\psi_3)^{T}$ and
\begin{equation*}
U_2=\left[
\begin{matrix}
                   0 & \lambda m & 1\\
                   0 & 0 & \lambda m \\
                   1 & 0 & 0 \\
                  \end{matrix}
                \right],\quad V_2=\left[
\begin{matrix}
                  -uu_x & \frac{u_x}{\lambda}-\lambda u^2m  & u_x^2\\
                   \frac{u}{\lambda} & -\frac{1}{\lambda^2} & -\lambda u^2m-\frac{u_x}{\lambda} \\
                   -u^2 &  \frac{u}{\lambda} & uu_x \\
                  \end{matrix}
                \right].
\end{equation*}
In such a case, the reciprocal transformation (\ref{gxre}) reduces to
\begin{equation}\label{novre}
dy=p^2dx-u^2p^2dt,\quad \quad \quad d\tau=dt,
\end{equation}
with $p=m^{\frac{1}{3}}$. This is a reciprocal transformation of the Novikov equation which changes
 the Lax pair (\ref{novlax}) to
\begin{equation}\label{anovlax}
\psi_y=F_2\psi,\quad \quad \quad \psi_\tau=G_2\psi,
\end{equation}where
\begin{equation*}
F_2=\left[
\begin{matrix}
                  0 & \lambda p & \frac{1}{p^2}\\
                  0 & 0 & \lambda p \\
                 \frac{1}{p^2} & 0 & 0 \\
                  \end{matrix}
                \right],\quad\quad\quad G_2=\left[
\begin{matrix}
                -u_yup^2 & \frac{u_yp^2}{\lambda}  & u^2+u_y^2p^4\\
                   \frac{u}{\lambda} & -\frac{1}{\lambda^2} & -\frac{u_yp^2}{\lambda} \\
                  0 &   \frac{u}{\lambda} & u_yup^2 \\
                  \end{matrix}
                \right].
\end{equation*}The compatibility condition of (\ref{anovlax}) yields the associated Novikov (aNovikov) equation \cite{Hone,Rasin3}
\begin{equation}\label{anovikov}
p_\tau=-p^3uu_y,\quad \quad \quad u_{yy}p^4+2p^3p_yu_y+p^3-u=0.
\end{equation}
It is easy to show that the scalar spectral problem of Lax pair (\ref{anovlax}) with respect to $\psi_2$ is just that of the SK hierarchy
\begin{equation}
[\partial_y^3-(\frac{p_{yy}}{p}+\frac{1}{p^4})\partial_y]\psi_2=\lambda^2\psi_2.
\end{equation}
With the help of DT for the SK hierarchy \cite{Leble}, the following Proposition holds.
\begin{prop}
The Lax pair (\ref{anovlax}) is covariant with respect to the DT
\begin{equation}\label{novdt}
\begin{array}{l}
\psi_{[1]}=T\psi, \quad T={\rm diag}[\frac{p_{[1]}}{p},1,\frac{p}{p_{[1]}}]((\lambda^2+\lambda_1^2)I-\left[
\begin{matrix}
                  T_{11} & T_{12} & T_{13}\\
                  \frac{2\lambda\lambda_1}{\sigma_2} & 2\lambda_1^2 & -2\lambda\lambda_1\frac{\sigma_1}{\sigma_2} \\
                 -2\frac{\lambda_1^2}{\sigma_2^2} & 2\frac{\lambda\lambda_1}{\sigma_2} & 2\frac{\lambda_1^2\sigma_1}{\sigma_2^2} \\
                  \end{matrix}
                \right])\\
p^2_{[1]}=p^2|(1-2\frac{\sigma_1}{\sigma_2^2})^2-\frac{4}{\sigma_2^4}|, \\
u_{[1]}=-\frac{p}{p_{[1]}}(u+\frac{2p^2u_y-2u\sigma_1}{\sigma_2^2}+\frac{2}{\lambda_1\sigma_2}).
\end{array}
\end{equation}
where
$$\begin{array}{l}
T_{11}=2\frac{\lambda_1^2}{\sigma_2^2}(p^2\frac{p_{[1],y}}{p_{[1]}}-pp_y-\sigma_1)+4p^3\frac{\lambda_1^3}{\sigma_2^3}, \nonumber\\
T_{12}=2\frac{\lambda\lambda_1}{\sigma_2}(\sigma_1-p^2\frac{p_{[1],y}}{p_{[1]}}+pp_y)-\frac{4\lambda\lambda_1^2p^3}{\sigma_2^2},\\
T_{13}=(\lambda^2+\lambda_1^2-2\frac{\lambda_1^2\sigma_1}{\sigma_2^2})(2\frac{\lambda_1p^3}{\sigma_2}+p^2\frac{p_{[1],y}}{p_{[1]}}-pp_y)
+2\frac{\lambda_1^2\sigma_1^2}{\sigma_2^2},\nonumber
\end{array}$$
 and $\sigma_1=\frac{g_1}{g_3},\sigma_2=\frac{g_2}{g_3}$,  $(g_1,g_2,g_3)^{T}$ is a special solution of the linear system (\ref{anovlax}) at $\lambda=\lambda_1$.
\end{prop}

Now, we shall establish a BT for the Novikov equation with the help of reciprocal transformation (\ref{novre}).
It infers from the Proposition 5 that
\begin{equation}
p^2_{[1]}=\frac{4p^2}{\sigma_2^4}|\vartheta^2-1|, \quad\quad\quad  \vartheta=\frac{1}{2}\sigma_2^2-\sigma_1.
\end{equation}
If $\vartheta^2-1>0$, direct calculation shows that
\begin{equation}\label{novx}
\frac{1}{p^2_{[1]}}=\frac{1}{p^2}-\frac{\vartheta_y}{\vartheta^2-1},\quad \quad \quad u^2_{[1]}=u^{2}-\frac{\vartheta_\tau}{\vartheta^2-1},
\end{equation}
Substituting (\ref{novx}) into (\ref{novre}) and taking the integration constant to be zero, we obtain
\begin{equation}\label{nox}
x_{[1]}=x+\frac{1}{2}{\rm ln} \lvert \frac{\vartheta+1}{\vartheta-1}\rvert.
\end{equation}
If $\vartheta^2-1<0$, a similar process gives rise to
\begin{equation}
x_{[1]}=-x-\frac{1}{2}{\rm ln} \lvert \frac{\vartheta+1}{\vartheta-1}\rvert.
\end{equation}
\begin{cor}
A BT of the Novikov equation reads
\begin{eqnarray}\label{novbt}
\begin{aligned}
&x_{[1]}=x+\frac{1}{2}{\rm ln} \lvert \frac{\vartheta+1}{\vartheta-1}\rvert,\quad\quad\quad t_{[1]}=t,\\
&u_{[1]}=\pm\frac{1}{\sqrt{\vartheta^2-1}}(u\vartheta+u_x+\frac{\eta}{\lambda_1}),
\end{aligned}\end{eqnarray}
if $\vartheta^2-1>0$, and
\begin{eqnarray}\label{novbt2}
\begin{aligned}
&x_{[1]}=-x-\frac{1}{2}{\rm ln} \lvert \frac{\vartheta+1}{\vartheta-1}\rvert,\quad\quad\quad t_{[1]}=t,\\
&u_{[1]}=\pm\frac{1}{\sqrt{1-\vartheta^2}}(u\vartheta+u_x+\frac{\eta}{\lambda_1}),
\end{aligned}\end{eqnarray}
if $\vartheta^2-1<0$. Here $\theta=\frac{1}{2}\eta^2+\frac{\eta_x}{\eta}-\lambda_1\frac{m}{\eta}$ and $\eta$ is determined by
\begin{eqnarray*}
&&\eta_{xx}=\lambda_1m_x-\eta-\lambda_1m\eta^2+(2\eta_x^2-3\lambda_1m\eta_x+\lambda_1^2m^2)/\eta, \\
&&\eta_t=-(u^2+\frac{u}{\lambda_1\eta})\eta_x-\eta(uu_x+\frac{1}{\lambda_1^2})-\frac{u_x}{\lambda_1}+\frac{um}{\eta}-\frac{u\eta^2}{\lambda_1}.
\end{eqnarray*}
\end{cor}
{\bf Remark:} Comparing the BT (\ref{novbt}) with the one in \cite{mh}, we may show that they are related by  $a=\frac{2\lambda_1p}{\sigma_2}$.

\section{Appendix: proofs of the identities (\ref{plu2}) and (\ref{jac2})}
In this section, we will give proofs of the identities \eqref{jac}, \eqref{plu2} and \eqref{jac2}. Actually, the \eqref{jac} is the direct result of the Jacobi identity. Since proofs of identities in \eqref{plu2} are similar, we only prove one of them, and so do \eqref{jac2}. Here we prove the first one in \eqref{plu2} for $N=3k+2$.
For convenience, we define
\begin{equation}\label{a}
\vec{\alpha}_N=(\alpha_1,...,\alpha_N)^{T},\quad \lambda^k\vec{\alpha}_N=(\lambda^k_1\alpha_1,...,\lambda^k_N\alpha_N)^{T},\quad \vec{1}_N=(1,...,1)^{T},
\end{equation}and $\vec{1}_N(i)$ as the column vector with the $i$-th element $1$ and other elements $0$.
Then, using the Pl\"ucker relation,  we compute
\begin{eqnarray*}
&&\hspace{-0.3cm}A_{N-1}B_N-A_NB_{N-1}\\
&&=\left|\begin{matrix}
                                        \vec{a}_{N-1} & \lambda \vec{b}_{N-1} &  \cdots & \lambda^{2k}\vec{a}_{N-1} \\
                                     \end{matrix}\right| \left|\begin{matrix}
                                        \vec{1}_{N} & \lambda\vec{b}_{N} &  \cdots & \lambda^{2k+1}\vec{b}_{N} \\
                                     \end{matrix}\right|\\
&&\hspace{0.4cm}-\left|\begin{matrix}
                                        \vec{a}_{N} & \lambda\vec{b}_{N} &  \cdots & \lambda^{2k+1}\vec{b}_{N} \\
                                     \end{matrix}\right|\left|\begin{matrix}
                                        \vec{1}_{N-1} & \lambda \vec{b}_{N-1} &  \cdots & \lambda^{2k}\vec{a}_{N-1} \\
                                     \end{matrix}\right| \\
&&=\left|\begin{matrix}
                                        \vec{a}_{N} & \lambda\vec{b}_{N} &  \cdots & \lambda^{2k}\vec{a}_{N} &\vec{1}_{N}(N)\\
                                     \end{matrix}\right|\left|\begin{matrix}
                                        \vec{1}_{N} & \lambda\vec{b}_{N} &  \cdots & \lambda^{2k+1}\vec{b}_{N} \\
                                     \end{matrix}\right| \\
&&\hspace{0.4cm}-\left|\begin{matrix}
                                        \vec{a}_{N} & \lambda\vec{b}_{N} &  \cdots & \lambda^{2k+1}\vec{b}_{N} \\
                                     \end{matrix}\right|\left|\begin{matrix}
                                        \vec{1}_{N} & \lambda \vec{b}_{N} &  \cdots & \lambda^{2k}\vec{a}_{N} &\vec{1}_{N}(N) \\
                                     \end{matrix}\right|\\
&&= \left|\begin{matrix}
                                       \vec{a}_{N} & \lambda\vec{b}_{N} &  \cdots & \lambda^{2k}\vec{a}_{N} & \vec{1}_{N}\\
                                     \end{matrix}\right|\left|\begin{matrix}
                                       \vec{1}_{N}(N)  &\lambda \vec{b}_{N} &  \cdots & \lambda^{2k}\vec{a}_{N}& \lambda^{2k+1}\vec{b}_{N}\\
                                     \end{matrix}\right|\\
&&=\Delta_N \Delta_{N+1}\left[\begin{array}{cc}
           N&N+1 \\ 1&2
         \end{array}
       \right].
\end{eqnarray*}

Now, we proceed to prove the first one of \eqref{jac2} as $N=3k+2$. With the aid of the Jacobi identity, we have
\begin{eqnarray*}
&&\hspace{-0.2cm}R_3=\lambda_{N+1}^2 \left|\begin{matrix}
                                        \vec{1}_{N-2} &\vec{a}_{N-2}  &  \cdots & \lambda^{2k}\vec{1}_{N-2} \\
                                        1 & a_{N+1}  &  \cdots & \lambda^{2k}_{N+1}\\
                                     \end{matrix}\right|\left|\begin{matrix}
                                         \lambda\vec{b}_{N-1} &  \cdots & \lambda^{2k+1}\vec{b}_{N-1} \\
                                     \end{matrix}\right|\\
&&\hspace{0.8cm}-\lambda_{N-1}^2 \left|\begin{matrix}
                                        \vec{1}_{N-1} &  \vec{a}_{N-1} &  \cdots & \lambda^{2k}\vec{1}_{N-1} \\
                                     \end{matrix}\right|\left|\begin{matrix}
                                         \lambda\vec{b}_{N-2} &  \cdots & \lambda^{2k+1}\vec{b}_{N-2} \\
                                         \lambda_{N+1}b_{N+1}& \cdots & \lambda^{2k+1}_{N+1}b_{N+1}
                                     \end{matrix}\right|\\
&&\hspace{0.4cm}=\prod\limits_{i=1}^{N-2}\lambda_i^{-2}\{\left|\begin{matrix}
                                        \lambda^2\vec{1}_{N-2}  &  \cdots & \lambda^{2k+2}\vec{1}_{N-2} \\
                                        \lambda_{N+1}^2&  \cdots & \lambda_{N+1}^{2k+2}
                                     \end{matrix}\right|\left|\begin{matrix}
                                         \lambda\vec{b}_{N-1} &  \cdots & \lambda^{2k+1}\vec{b}_{N-1} \\
                                     \end{matrix}\right|     \\
&&\hspace{0.8cm}-\left|\begin{matrix}
                                        \lambda^2\vec{1}_{N-1}  &  \cdots & \lambda^{2k+2}\vec{1}_{N-1} \\
                                     \end{matrix}\right|\left|\begin{matrix}
                                         \lambda\vec{b}_{N-2} &  \cdots & \lambda^{2k+1}\vec{b}_{N-2} \\
                                         \lambda_{N+1}b_{N+1}& \cdots & \lambda^{2k+1}_{N+1}b_{N+1}
                                     \end{matrix}\right|\}\\
&&\hspace{0.4cm}=\prod\limits_{i=1}^{N-2}\lambda_i^{-2}\{\triangle_{N+2}\left[
\begin{matrix}
                 N-1 & N& N+2 \\
               1 & 2 & 3\\
                  \end{matrix} \right]\triangle_{N+1}\left[
\begin{matrix}
                 N & N+1 \\
               1 & 2 \\
                  \end{matrix} \right]\\
&&\hspace{0.8cm}-\triangle_{N+2}\left[
\begin{matrix}
                 N & N+1& N+2 \\
               1 & 2 & 3\\
                  \end{matrix} \right]\triangle_{N+1}\left[
\begin{matrix}
                 N-1 & N \\
               1 & 2 \\
                  \end{matrix} \right]\} \\
&&\hspace{0.4cm}=\prod\limits_{i=1}^{N-2}\lambda_i^{-2}\{\triangle_{N+2}\left[
\begin{matrix}
                 N & N+2& N-1 \\
               1 & 2 & 3\\
                  \end{matrix} \right]\triangle_{N+2}\left[
\begin{matrix}
                 N & N+2 &N+1 \\
               1 & 2 &N+2 \\
                  \end{matrix} \right]\\
&&\hspace{0.8cm}-\triangle_{N+2}\left[
\begin{matrix}
                 N & N+2& N+1 \\
               1 & 2 & 3\\
                  \end{matrix} \right]\triangle_{N+2}\left[
\begin{matrix}
                N&  N+2 & N-1 \\
               1 & 2 & N+2\\
                  \end{matrix} \right]\}    \\
\end{eqnarray*}

\begin{eqnarray*}
&&\hspace{0.4cm}=\prod\limits_{i=1}^{N-2}\lambda_i^{-2}\triangle_{N+2}\left[
\begin{matrix}
                 N& N+2 \\
               1 & 2 \\
                  \end{matrix} \right]\triangle_{N+1}\left[
\begin{matrix}
                 N-1& N & N+1 \\
               1 & 2 &3\\
                  \end{matrix} \right].
\end{eqnarray*}

\bigskip
\noindent
\section*{Acknowledgements}
This work is partially supported by the National Natural Science Foundation of China (Grant Nos. 12271190 and 11871232), and Youth Innovation Foundation of Xiamen (project no. 3502Z20206011),


\section*{References}
\begin{thebibliography}{20}
\bibitem{Holm}R. Camassa and D. D. Holm, An integrable shallow water equation with peaked solitons, Phys. Rev. Lett. {\bf 71} (1993) 1661-1664.
\bibitem{Rasin}A. G. Rasin and J. Schiff, B\"acklund transformations for the Camassa-Holm equation, J. Nonlinear Sci. {\bf 27} (2017) 45-69.
\bibitem{Constantin} A. Constantin, On the scattering problem for the Camassa-Holm equation, Proc. R. Soc. London, Ser. A {\bf 457} (2001) 953-970.
\bibitem{Gerdjikov}A. Constantin, V. S. Gerdjikov and R. I. Ivanov, Inverse scattering transform for the Camassa-Holm equation, Inverse Problems {\bf 22} (2006) 2197-2207.
\bibitem{Fuchssteiner}B. Fuchssteiner, Some tricks from the symmetry-toolbox for nonlinear equations: generalizations of the Casmassa-Holm equation, Phys. D {\bf 95} (1996) 229-243.
\bibitem{Szmigielski}R. Beals, D. H. Sattinger and J. Szmigielski, Multi-peakons and a theorem of Stieltjes, Inverse Problems  {\bf 15} (1999)
L1-4.
\bibitem{Sattinger} R. Beals, D. H. Sattinger and J. Szmigielski,  Multipeakons and the classical moment problem, Adv. Math.  {\bf 154} (2000)
229-257.
\bibitem{Beals}R. Beals, D. H. Sattinger and J. Szmigielski, Peakons, strings, and the finite Toda lattice, Comm. Pure Appl. Math. {\bf 54} (2001) 91-106.              
\bibitem{Popowicz}N. Li, Q. P. Liu and Z. Popowicz, A four-component Camassa-Holm type hierarchy, J. Geom. Phys. {\bf 85} (2014) 29-39.
\bibitem{Geng}X.  Geng and B. Xue, An extension of integrable peakon equations with cubic nonlinearity, Nonlinearity {\bf 22} (2009)  1847-1856.
\bibitem{Liuq}N. Li and Q. P. Liu, On bi-Hamiltonian structure of two-component Novikov equation, Phys. Lett. A {\bf 377} (2013) 257-261.
\bibitem{Li}N. Li and X. Niu, A reciprocal transformation for the Geng-Xue equation, J. Math. Phys. {\bf 55} (2014) 053505.
\bibitem{ldk}H. Lundmark and J. Szmigielski, An inverse spectral problem related to the Geng-Xue two-component peakon equation, Mem. Amer. Math. Soc.     {\bf 244} (2016) viii+87 pages.
\bibitem{Liliu}N. Li and Q. P. Liu, Smooth multisoliton solutions of a 2-component peakon
system with cubic nonlinearity, SIGMA {\bf 18} (2022), 066.
\bibitem{Degasperis}A. Degasperis and M. Procesi, Asymptotic integrability in Symmetry and Perturbation Theory edited by A. Degasperis and G. Gaeta, World Scientific, Singapore, 1999, pp 23-37.
\bibitem{Novikov}V. Novikov,  Generalizations of the Camassa-Holm equation, J. Phys. A: Math. Theor. {\bf 42} (2009) 342002.
\bibitem{wang1}A. N. W. Hone  and J. P. Wang, Prolongation algebras and Hamiltonian operators for peakon equations, Inverse Problems {\bf 19} (2003) 129-145.
\bibitem{Hone}A. N. W. Hone and J. P. Wang, Integrable peakon equations with cubic nonlinearity,  J. Phys. A: Math. Theor. {\bf 41} (2008) 372002.
\bibitem{Deg2}A. Degasperis, D. D. Holm and A. N. W. Hone, A new integrable equation with peakon solitons, Theor. Math. Phys. {\bf 133} (2002) 1463-1474.
\bibitem{Matsuno1}Y. Matsuno, Multisoliton solutions of the Degasperis-Procesi equation and their peakon limit, Inverse problems,  {\bf 21} (2005) 1553-1570.
\bibitem{Matsuno2}Y. Matsuno, Smooth multisoliton solutions and their peakon limit of Novikov's Camassa-Holm type equation with cubic nonlinearity, J. Phys. A: Math. Theor. {\bf 46} (2013) 365203.
\bibitem{Li2}N. Li, G. Wang and Y. Kuang, Multisoliton solutions of the Degasperis-Procesi
equation and its short-wave limit: Darboux transformation approach, Theor. Math. Phys. {\bf 203} (2020) 608-620.
\bibitem{Wu}L. Wu, C. Li and N. Li, Soliton solutions to the Novikov equation and a negative flow of
the Novikov hierarchy, Appl. Math. Lett. {\bf 87} (2019) 134-140.

\bibitem{Ivco}A. Constantin and R. Ivanov, Dressing method for the Degasperis-Procesi equation, Stud. Appl. Math. {\bf 138} (2017) 205-226.

\bibitem{Lundmark}A. N. W. Hone, H. Lundmark  and J. Szmigielski, Explicit multipeakon solutions of Novikov's
cubically nonlinear integrable Camassa-Holm type equation, Dyn. PDE {\bf 6} (2009) 253-289.
\bibitem{Szmigielski2}H. Lundmark  and J. Szmigielski, Multi-peakon solutions of the Degasperis-Procesi
equation, Inverse Problems {\bf 19} (2003) 1241-1245.
\bibitem{Szmigielski3}H. Lundmark and J. Szmigielski, Degasperis-Procesi peakons and the discrete cubic string, Int.
Math. Res. Pap. {\bf 2005} (2005) 53-116.
\bibitem{b}  A.V. B\"acklund A V. Zur theorie der Fl\"achentransformationen, Math. Ann. 1881, {\bf 19} (1881) 387-422.
\bibitem{la} R. M. Miura, B\"acklund transformations, the inverse scattering method, solitons, and their applications. Springer, New York,
1976.
\bibitem{rs}C. Rogers, W.K. Schief. B\"acklund and Darboux transformations: geometry and modern applications in soliton
theory. Cambridge University Press, Cambridge, 2002.

\bibitem{mw}H. Mao and G. Wang, B\"acklund Transformations for the Degasperis-Procesi Equation, Theor. Math. Phys. {\bf 203} (2020) 747-760.
\bibitem{mh}H. Mao,  Novikov equation: B\"acklund transformation and applications, Theor. Math. Phys. {\bf 206} (2021) 163-173.
\bibitem{ml}H. Mao, and Q. P. Liu, The short pulse equation: B\"{a}cklund transformations and applications, Stud. Appl. Marh. {\bf145} (2020) 791-811.
\bibitem{Huang}S. Huang and H. Li, Darboux transformations of the Camassa-Holm type systems, Chaos, Solitons and Fractals {\bf 157} (2022) 111910.
\bibitem{Leble} S. B. Leble and N. V. Ustinov, Third order spectral problems: reductions and Darboux
transformations, Inverse Problems {\bf 10} (1994) 617-633.
\bibitem{Rasin2}A. G. Rasin and J. Schiff, Unfamiliar aspects of B\"acklund transformations and an associated Degasperis-Procesi equation, Theor. Math.  Phys. {\bf 196} (2018) 1333-1346.
\bibitem{Rasin3}A. G. Rasin and J. Schiff, A simple-looking relative of the Novikov, Hirota-Satsuma and Sawada-Kotera equations, J. Nonlinear Math. Phys. {\bf 26} (2019) 555-568.

\end{thebibliography}
\end{document}